\documentclass[usenatbib]{mnras}

\usepackage{graphicx}	
\usepackage{array}
\usepackage{amsmath}	
\usepackage{footnote}

\newcommand{\degree}{\ensuremath{^\circ}}

\title[Infrared color and variability of NLSy1 galaxies]{{\it WISE} view of Narrow-Line Seyfert 1 galaxies: mid-infrared color and variability}

\author[Rakshit et al.]{Suvendu Rakshit$^{1,2}$\thanks{E-mail: suvenduat@gmail.com},
Ansu Johnson$^{2,3}$\thanks{Project work carried out at IIA as part of the IASc-INSA-NASI Summer Research Fellowship -2017},
C. S. Stalin$^{2}$,
Poshak Gandhi$^{4}$,
\newauthor 
Sebastian Hoenig$^{4}$
\\\\
$^{1}$Astronomy Program, Department of Physics and Astronomy, Seoul National University, Seoul 151-742, Republic of Korea \\
$^{2}$Indian Institute of Astrophysics, Block II, Koramangala, Bangalore 560 034, India\\
$^{3}$National Institute of Technology Tiruchirappalli - 620015 \\
$^{4}$Department of Physics \& Astronomy, University of Southampton, Southampton, SO17 1BJ, UK\\
}

\date{}

\pubyear{2018}

\begin{document}
\label{firstpage}
\pagerange{\pageref{firstpage}--\pageref{lastpage}}
\maketitle


\begin{abstract}
We present the color and flux variability analysis at 3.4 $\mu m$ ($W$1-band) and 4.6 $\mu m$ ($W$2-band) of 492 narrow-line Seyfert 1 (NLSy1) galaxies using archival data from the Wide-field Infrared Survey Explorer ({\it WISE}). In the {\it WISE} color-color, ($W1 - W2$) versus ($W2-W3$) diagram, $\sim 58\%$ of the NLSy1 galaxies of our sample lie in the region occupied by the blazar category of active 
galactic nuclei (AGN). The mean $W1-W2$ color of candidate variable NLSy1 galaxies is $0.99 \pm 0.18$ mag. The average amplitude of variability is $0.11 \pm 0.07$ mag in long-term (multi-year) with no difference in variability between
$W$1 and $W$2-bands. The $W1-W2$ color of NLSy1 galaxies is anti-correlated with the relative strength of [O III] to H$\beta$, strongly 
correlated with continuum luminosity, black hole mass, and Eddington ratio. 
The long-term amplitude of variability shows weak anti-correlation with the Fe II 
strength, continuum luminosity and Eddington ratio. A positive correlation between color as well as 
the amplitude of variability with the radio power at 1.4 GHz was found for the 
radio-detected NLSy1 galaxies. This suggests non-thermal synchrotron 
contribution to the mid-infrared color and flux variability in radio-detected NLSy1 galaxies.
\end{abstract}




\section{Introduction}

Active Galactic Nuclei (AGN) are the persistent luminous objects in the sky
powered by accretion of matter onto the super-massive black holes (SMBHs)
located at the centers of galaxies \citep{1969Natur.223..690L,1984ARA&A..22..471R}.
Models of AGN, in addition to the central SMBH, posit the presence of accretion disc surrounded by the dusty obscuring torus \citep{1993ARA&A..31..473A,1995PASP..107..803U}, which are supported by the observed big blue bump (BBB) and mid-infrared (mid-IR) bump in their broad band spectral energy distribution (SED).
The BBB is thought to be thermal emission from the accretion disk
\citep{1973A&A....24..337S} and the IR bump is from the dusty torus known
from dust reverberation mapping studies \citep{2006ApJ...639...46S, 2014ApJ...788..159K,
2018MNRAS.475.5330M}.  One of the important characteristics of AGN known
since their discovery is that they show flux variability
\citep{1997ARA&A..35..445U, 1995ARA&A..33..163W}. This
property of AGN,  observed at different wavelengths from low energy radio
to high energy $\gamma$-ray and on a broad range of timescales
from hours to years \citep[e.g.,][]{1995ARA&A..33..163W,1997ARA&A..35..445U,1999MNRAS.306..637G,2009ApJ...698..895K,2010ApJ...721.1014M,2011A&A...525A..37M,2017ApJ...842...96R,2018ApJ...854..160R,2018arXiv180507747L},
has been used to probe their central regions which
are beyond the reach of any imaging techniques \citep{2015MNRAS.447.2420R}. Optical variability
studies on large samples of AGN found interesting relations of
the amplitude and timescale of variability with the physical parameters of AGN
such as black hole mass, FeII strength, redshift etc.
\citep[e.g.,][]{1996ApJ...463..466D,2004ApJ...601..692V,2009ApJ...698..895K,2010ApJ...721.1014M,2011A&A...525A..37M,2017ApJ...842...96R,2018arXiv180507747L} providing new insights into the accretion process in AGN.

Various types of AGN are known and one among them are the narrow line
Seyfert 1 (NLSy1) galaxies, which are classified based on the presence of
narrow H$\beta$ emission line with full width at half maximum (FWHM)
$<2000$ km s$^{-1}$ and weak [O III] emission line, with
F([O III])/F(H$\beta$) $<3$ \citep{1985ApJ...297..166O,1989ApJ...342..224G}.
They are believed to be powered by low mass black holes ($\sim 10^7 M_{\odot}$)
having higher accretion rate and generally showing strong Fe II emission
compared to their broad line counterparts namely the broad-line Seyfert 1 (BLSy1)
galaxies \citep{2012AJ....143...83X,2017ApJS..229...39R}. However,
from spectro-polarimetric observations of a $\gamma$-ray emitting
NLSy1 galaxy,  PKS 2004$-$447 \citep{2016MNRAS.458L..69B} and accretion disk modeling of a
sample of 23 radio-detected NLSy1 galaxies \citep{2013MNRAS.431..210C} indicate that they have masses similar to the blazar class of AGN.
 Other characteristics that make NLSy1 galaxies different from the BLSy1
galaxies are their rapid soft X-ray variability \citep{1999ApJS..125..297L,1995MNRAS.277L...5P}, steep soft X-ray spectra
\citep{1996A&A...305...53B,1996A&A...309...81W,1999ApJS..125..317L} and low
amplitude optical variability \citep{2004AJ....127.1799G,2017ApJ...842...96R}. Also,
the fraction of NLSy1 galaxies detected in
radio is much lower ($\sim 7\%$) compared to the fraction of radio detected
BLSy1 galaxies \citep{2006AJ....132..531K,2017ApJS..229...39R}. Among radio-loud
NLSy1 galaxies, about a dozen ($\sim$2\%) have been detected in $\gamma$-ray
by the {\it Fermi}-Large Area Telescope \citep[e.g.,][]{2009ApJ...699..976A,
2015MNRAS.452..520D,2018ApJ...853L...2P} suggesting the unambiguous presence of relativistic jets
in them. Multi-band broad band SED modeling of these $\gamma$-ray detected
NLSy1 galaxies  indicate
that these sources have many properties similar to the blazar class of
AGN \citep{2013ApJ...768...52P} and specifically resembling the flat spectrum radio quasar (FSRQ) category \citep{2018ApJ...853L...2P}. In the radio, these $\gamma$-ray emitting NLSy1 galaxies
have a compact core jet morphology, high brightness temperature, show
superluminal motion and significant radio variability 
\citep{2006AJ....132..531K,2006PASJ...58..829D}. Detailed
investigations of the population of NLSy1 galaxies need to be undertaken
to understand more about their peculiar characteristics.

The long-term (multi-year) optical variability nature of NLSy1 galaxies, compared
to their broad line counterparts  have been
recently investigated by \cite{2017ApJ...842...96R}, using a large number
of sources from the catalog of \cite{2017ApJS..229...39R} with the optical data taken
from the Catalina Real Time Transient Survey \citep[CRTS;][]{2009ApJ...696..870D}. In addition to the long-term optical variability, the intra-night
optical variability characteristics of different categories of NLSy1 galaxies
are also well studied \citep{2010ApJ...715L.113L,2013MNRAS.428.2450P,
2017MNRAS.466.2679K,2013ApJ...762..124M}. However, studies on the IR variability characteristics of NLSy1 galaxies are very limited. Rapid variability in the IR bands has been seen in three radio-loud NLSy1 galaxies on intra-day timescales \citep{2012ApJ...759L..31J}. Studies of IR variability are important as IR echoes in response to the optical continuum variations, can provide very crucial information about the dust torus morphology and the accretion mechanism \citep{2013ApJ...762..124M}. Moreover, IR wavelengths have some unique advantages over optical wavelengths, for example (1) IR continuum is unaffected by the presence of strong emission lines in low redshift AGN and (2) less affected by dust extinction. 

Given the limited studies on the IR variability
characteristics of NLSy1 galaxies, a systematic study on an unbiased sample
of sources is needed to give insight into their IR variability nature.
We, therefore, have carried out a systematic investigation
of the mid-IR flux variability characteristics of a sample of NLSy1 galaxies
taken from the catalog of \cite{2017ApJS..229...39R}. This is the first 
large-scale statistical study of the IR variability of NLSy1 galaxies both in
long-term as well as short-term (intra-day) 
containing few hundreds of
sources.  The sample and data used in this study are described in
Section \ref{sec:data},
our analysis is given in Section \ref{sec:results} and the summary of the
results are presented in Section \ref{sec:conclusion}. A cosmology with $H_0 = 70 \, \mathrm{km \, s^{-1} Mpc^{-1}}$, $\Omega_m = 0.3$, and $\Omega_{\lambda} = 0.7$ is assumed throughout and all magnitudes from WISE used here are in the Vega System.

\begin{figure}
\centering
\resizebox{8.5cm}{7.0cm}{\includegraphics{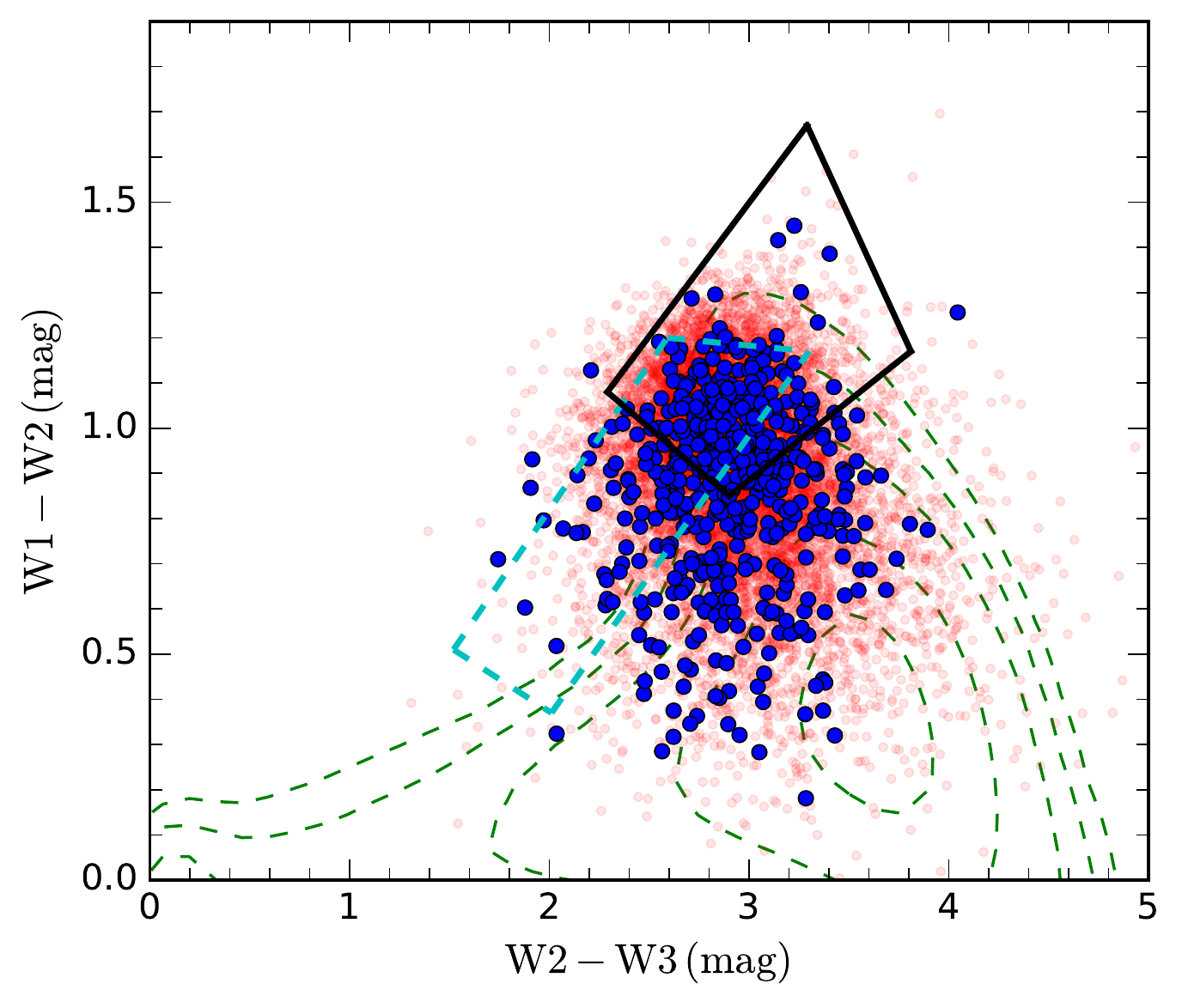}}
\caption{The {\it WISE} color-color diagram of the 520 variable candidates (blue circles) along with the parent NLSy1 galaxy sample (red dots). The {\it WISE} Gamma-ray strips for BL Lacs (dashed-cyan) and FSRQs (solid-black) is also shown. The green dashed contours represent {\it WISE} thermal sources (see text for explanation).}\label{Fig:color-color}. 
\end{figure}

\begin{figure}
\centering
\resizebox{8.5cm}{7.0cm}{\includegraphics{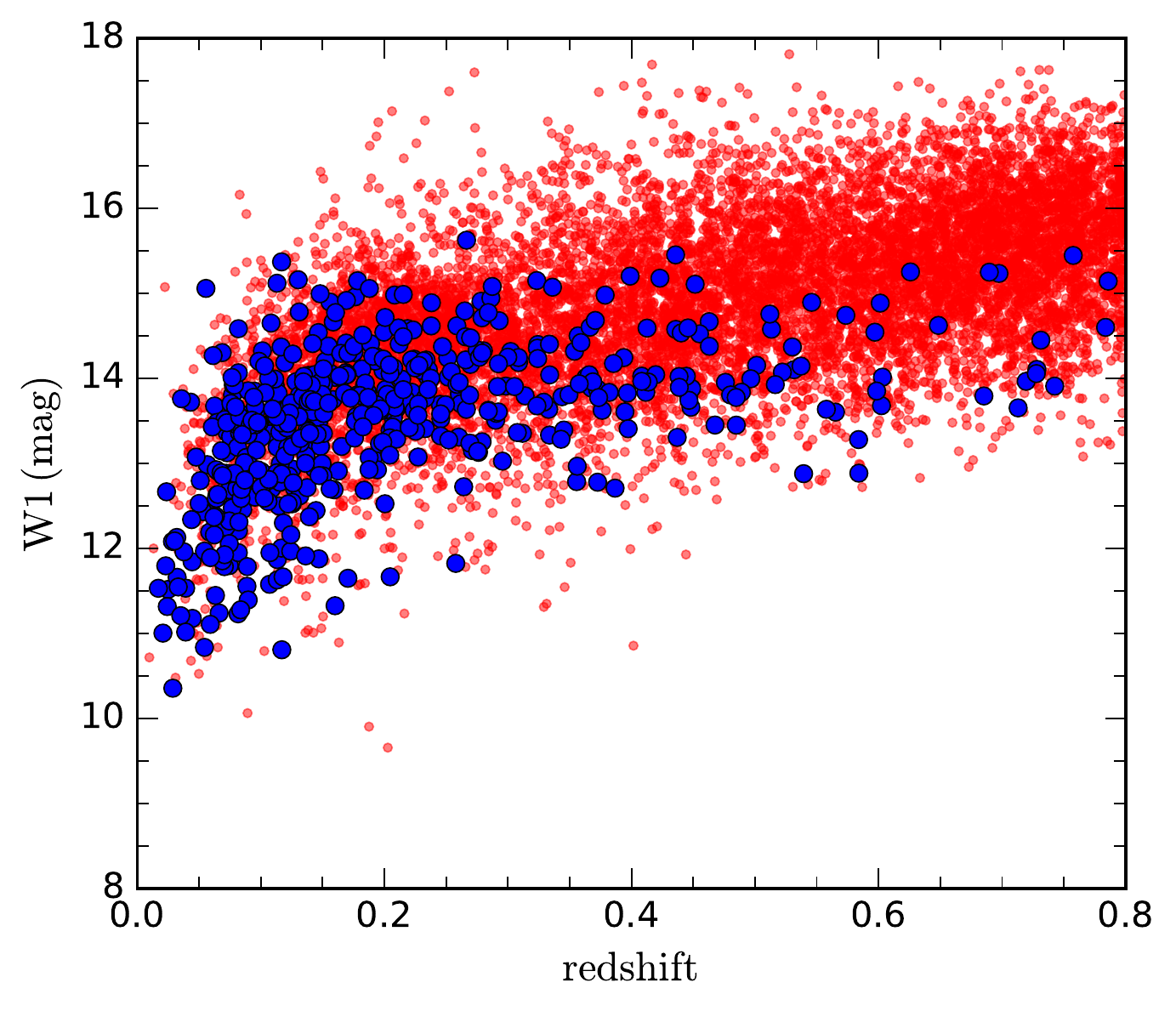}}
\caption{The $W$1 magnitude is plotted against the redshift of the 520 variable candidates (blue circles) along with the parent NLSy1 galaxy sample (red dots).}\label{Fig:redshift_magnitude}. 
\end{figure}

 \begin{figure*}
    \centering
    \resizebox{10.2cm}{5.5cm}{\includegraphics{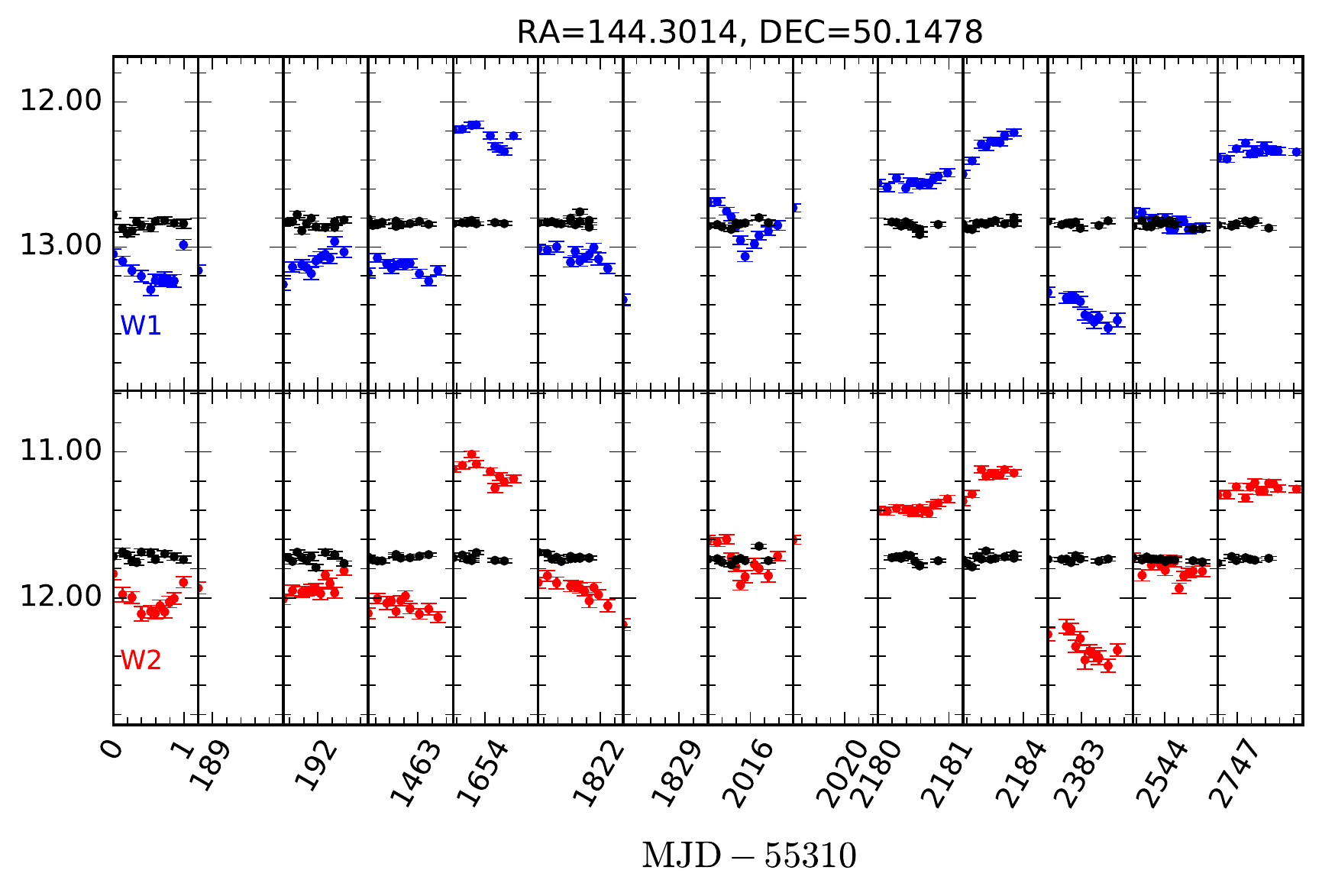}}
    \resizebox{5.4cm}{5.5cm}{\includegraphics{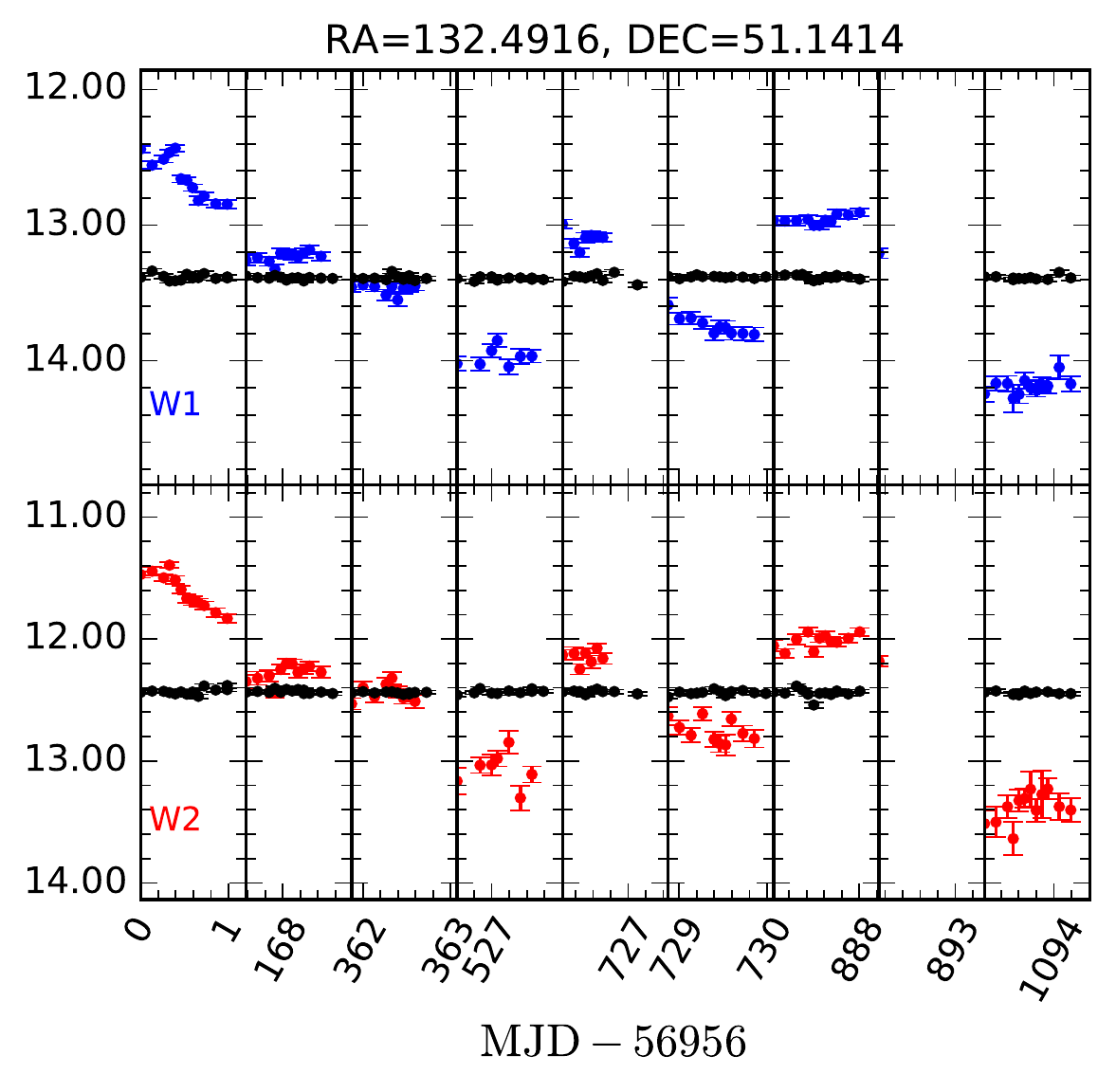}}
    \resizebox{8.1cm}{5.5cm}{\includegraphics{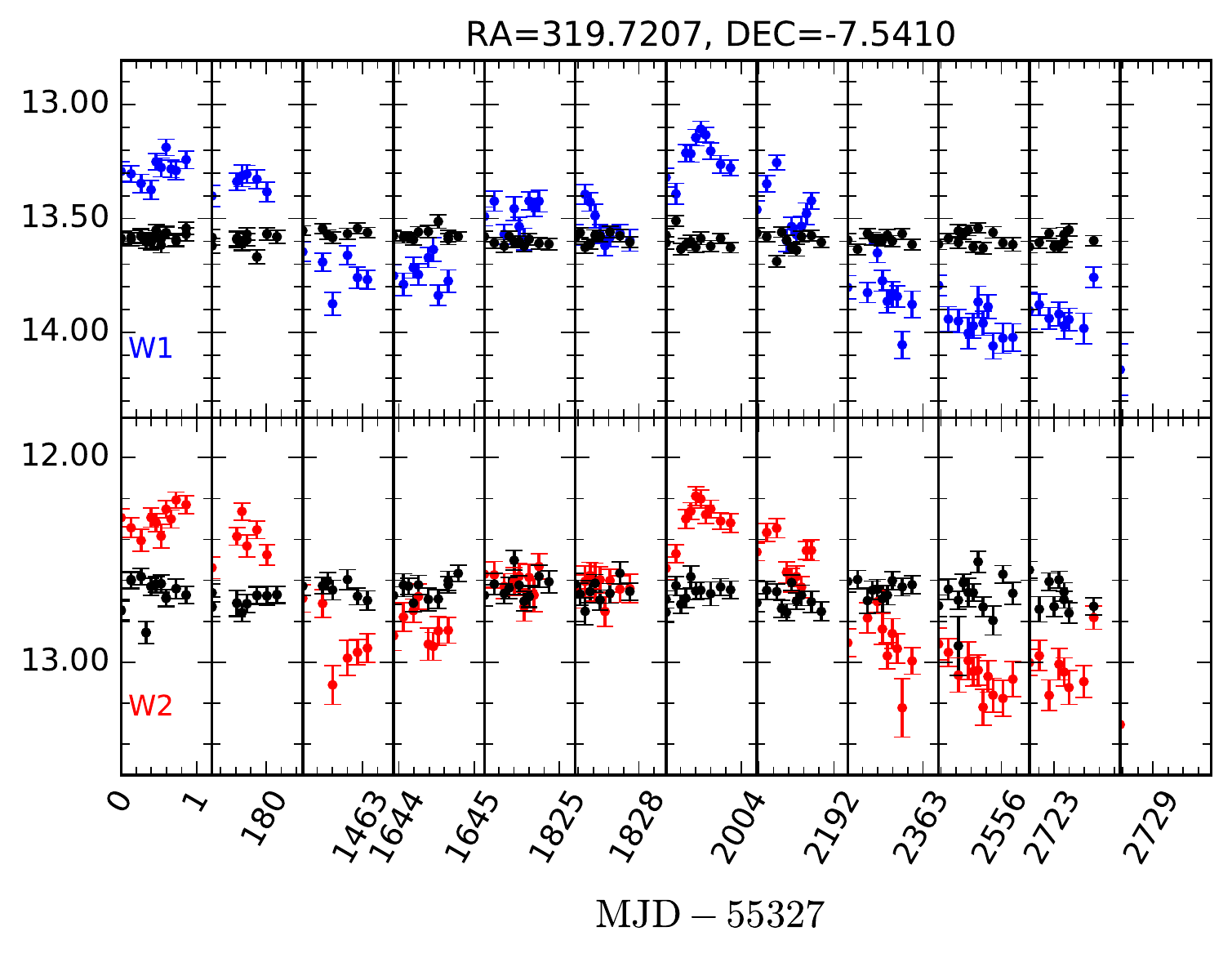}}
    \resizebox{8.1cm}{5.5cm}{\includegraphics{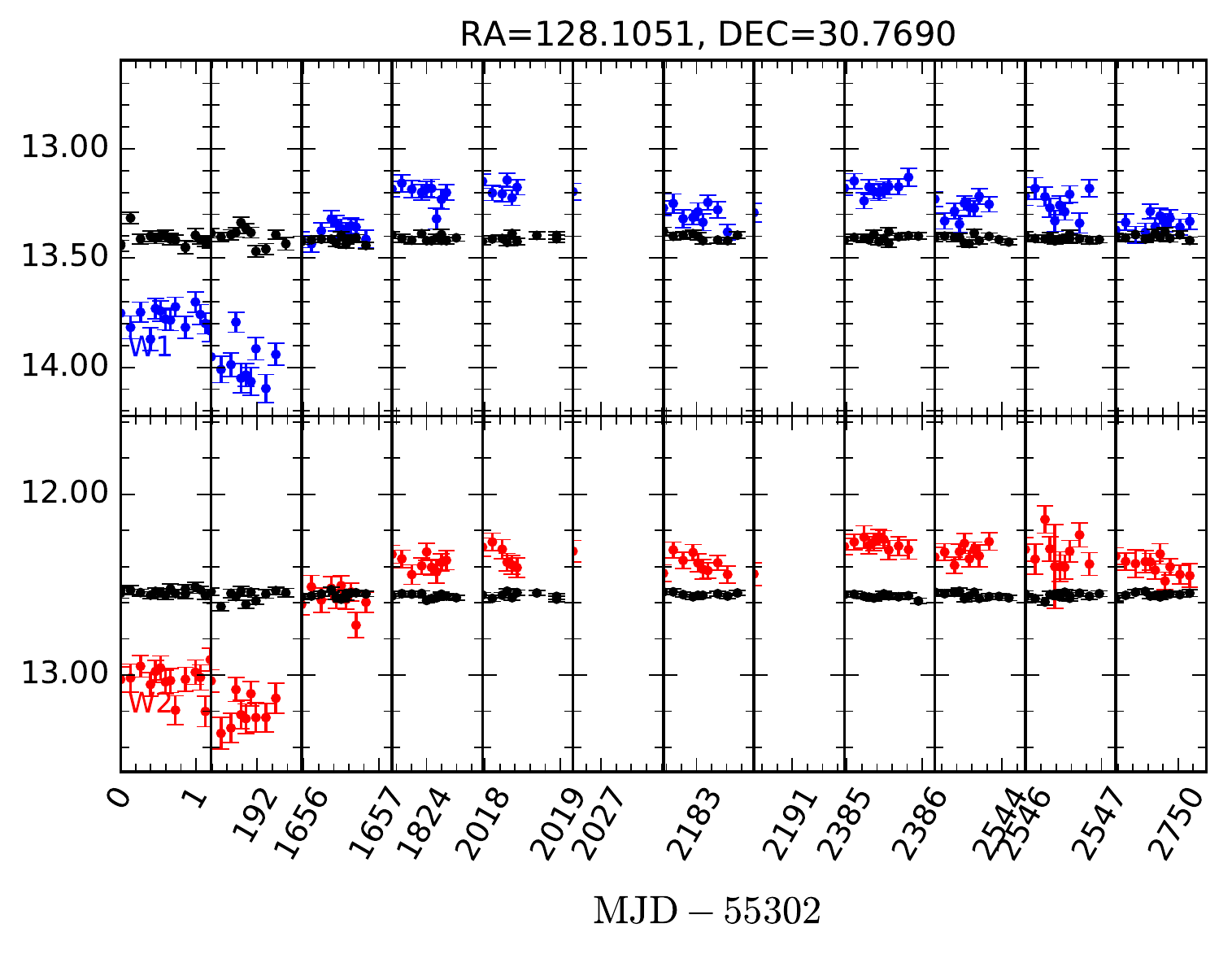}}
    \resizebox{8.1cm}{5.5cm}{\includegraphics{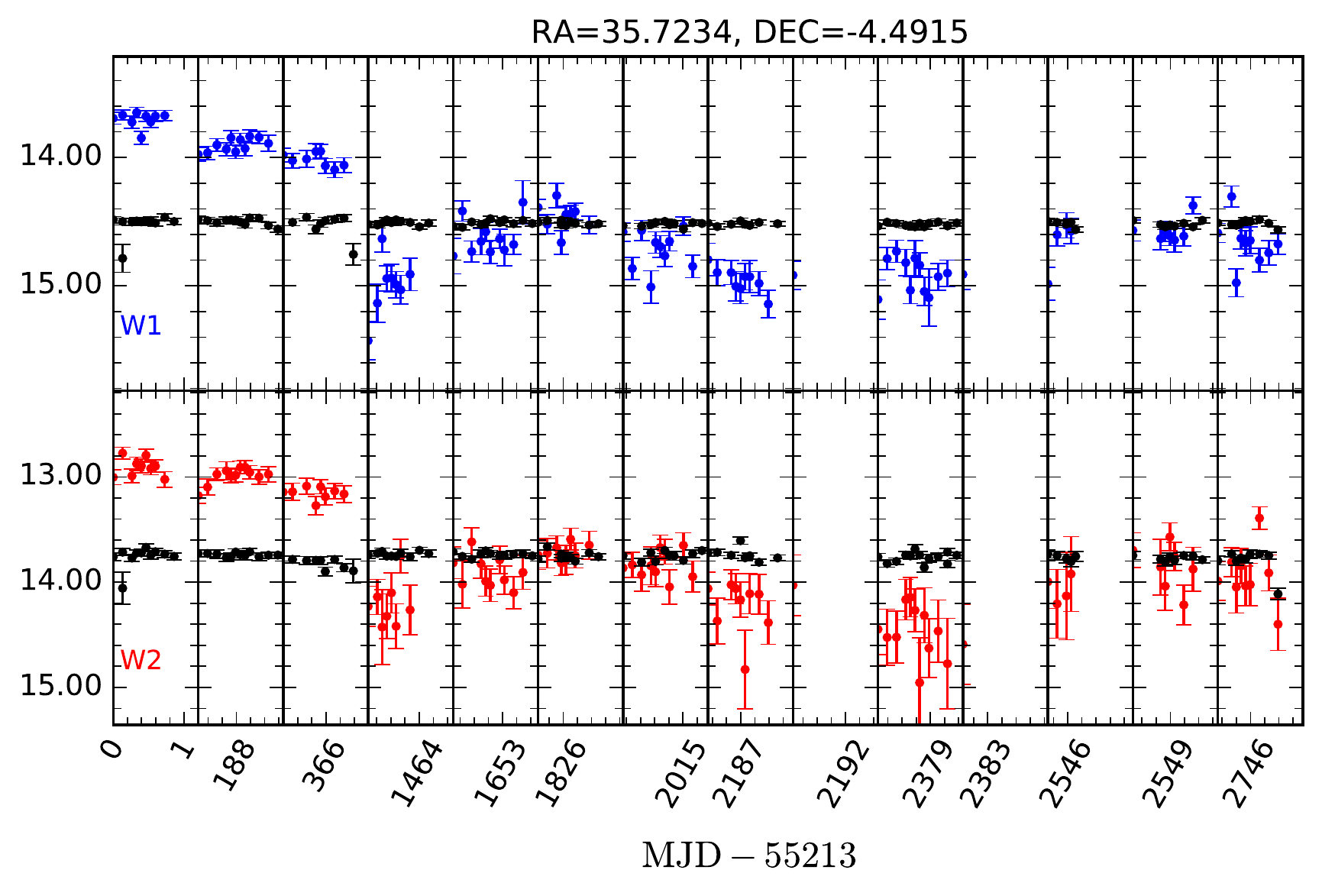}}
    \resizebox{8.1cm}{5.5cm}{\includegraphics{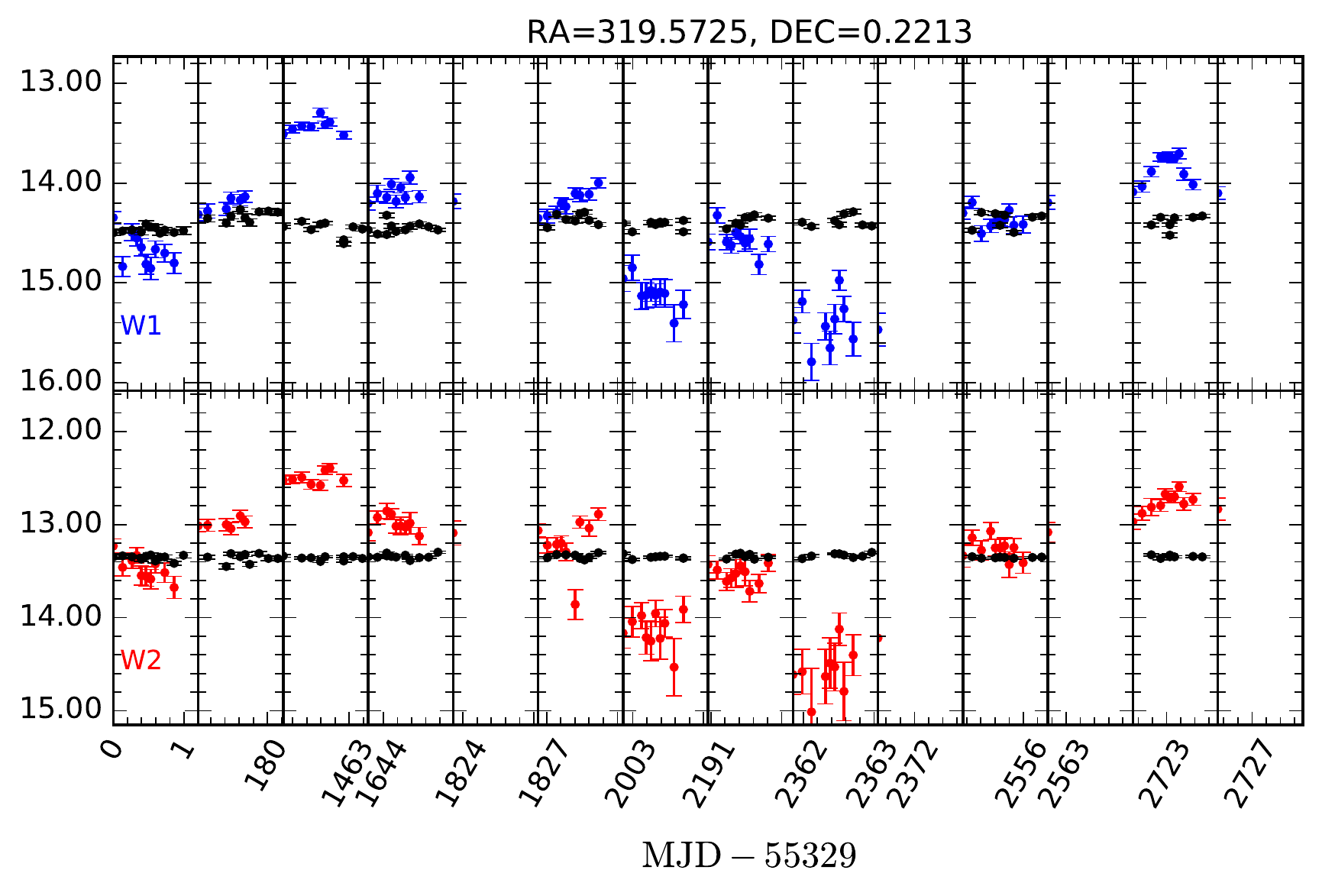}}
    
    \caption{Examples of {\it WISE} multi-epoch light curves in $W$1 (upper-panel) and $W$2 bands (lower-panel) for six NLSy1 galaxies (from top to bottom: 3FGL J0937.7+5008, 3FGL J0849.9+5108, SDSS J211852.96-073227.5, 2MASX J08322525+3046090, SDSS J022253.61-042929.2, 3FGL J2118.4+0013). The magnitudes of the objects in the Vega system are plotted along Y-axis. The light curve of a nearby star (magnitudes 8.4, 9.5, 12.2, 9.3, 11.6, 7.2 respectively from top to bottom in $W$1 band) present in the same field is also shown (black) after shifting by a constant value for comparison. The NLSy1 galaxies show large long-term variation compared to the stars. Each subplot has a size of 1.2 days.}\label{Fig:lc}. 
\end{figure*}

\section{Sample and data}\label{sec:data}
For this work, we have taken all NLSy1 galaxies from the catalog of 
\citet{2017ApJS..229...39R}, the selection of which was based on  the 
spectroscopic database of the Sloan Digital Sky Survey Data Release 12 
\citep[SDSS DR12;][]{2015ApJS..219...12A}. The catalog consists of 
a total of 11,101 NLSy1 galaxies. Since the main motivation of this work 
is to characterize the IR variability properties of the population of 
NLSy1 galaxies, we looked for the IR counterparts to these sources 
in the {\it Wide-field Infrared
Survey Explorer} \citep[{\it WISE};][]{2010AJ....140.1868W} database. {\it WISE} has mapped 
the entire sky in four different IR bands centered at 3.4 $\mu$m 
($W$1 band), 4.6 $\mu$m ($W$2 band), 12 $\mu$m ($W$3 band) and 22 $\mu$m 
($W$4 band) with angular 
resolution of 6.1, 6.4, 6.5 and 12 arcsec, respectively. It has much higher 
sensitivity than any previous all sky surveys in IR. {\it WISE} has a field of view of 47 arcmin $\times$ 47 arcmin and scans the entire sky once in every six months. The observations were performed (i) under WISE full cryogenic and NEOWISE Post-Cryo mission in three windows in 2010-2011 in $W$1, $W$2, $W$3 and $W$4 bands. All photometric magnitudes from this mission are included in the ``AllWISE MultiEpoch Photometry (MEP) Database'' and (ii) under the near-Earth objects {\it WISE} (NEOWISE) Reactivation mission \citep[hereafter, NEOWISE-R;][]{2014ApJ...792...30M} in 8 windows in 2014-2017 in $W$1 and $W$2 bands. All NEOWISE-R data are included in the ``NEOWISE-R Single Exposure (L1b) Source Table''. 

Within each visibility window, WISE observed sources continuously for a day, but sometimes more, with a cadence of 90 min, however, in some cases two exposures within 90 min were taken at 11s apart. The number of observations per window varied with ecliptic latitude. Thus, variability on short timescales of 90 minutes and long timescales of 6 months are most easily detected, but the observations are less
sensitive to other timescales. Such a temporal sampling is suitable enough for 
short-term and long-term variability studies. The ``AllWISE Source Catalog'' contains data from the {\it WISE}
cryogenic and NEOWISE post-cryogenic survey phases having about 747 million 
objects detected on the coadded atlas images. The catalog includes the variability 
flag, \texttt{var\textunderscore flg\footnote{For more information about \texttt{var\textunderscore flg} see \url{http://wise2.ipac.caltech.edu/docs/release/allwise/expsup/sec5_3bvi.html}}} from the observations done 
during 2010-2011. This flag, an integer 
value from 0 to 9, gives the probability of flux variation in each 
source. A value of 0 means no variability, while a value from 1 to 9
indicates increasing probability of variability \citep{2012AJ....143..118H}.

We cross-correlated all the 11,101 NLSy1 galaxies from \citet{2017ApJS..229...39R} 
with the ``AllWISE Source Catalog'' with a search radius of 2 arcsec centered 
at their optical position that resulted in 10,777 matches. From this
list of 10,777 sources, we imposed the criteria of the \texttt{var\textunderscore flg} equal to or greater than ``3'' since  \texttt{var\textunderscore flg} less than ``3'' can be considered as non-variable sources in the given band.  
The above criteria led us to a  sample of  520 objects.
The position of these sources (blue circles) in the color-color ($W$1-$W$2-$W$3) 
diagram is shown in Figure \ref{Fig:color-color}. Also in the same figure, our initial sample of 10,777 NLSy1 galaxies (red dots) having IR 
detections in {\it WISE} is shown.  The location of the {\it WISE} Gamma-ray strips 
\citep{2012ApJ...750..138M} for BL Lac and FSRQs
are shown by the cyan dashed and black solid lines, respectively. The green dashed lines \citep[see][]{2011wise.rept....1C,2011ApJ...740L..48M} are the density contours of 364,994 {\it WISE} thermal sources in a region of 48 
deg$^2$ area around high Galactic latitude ($l, b= 225 \degree, -55\degree$) 
taken from the {\it WISE} Preliminary Release Source Catalog (WPSC). Interestingly, $57.6\%$ of our sample of the 520 NLSy1 galaxies fall within 
the {\it WISE} Gamma-ray strips, of which 153 occupy the region 
populated both by BL Lacs and FSRQs. This similarity 
supports the detection of a handful of NLSy1 galaxies in $\gamma$-ray 
\citep{2009ApJ...699..976A,2018ApJ...853L...2P} but the detection might increase with the improvement in the sensitivity limit of future $\gamma$-ray instruments. Note that our sample of 520 sources has a  mean $W$1-$W$2 color of $0.99\pm 0.18$ mag. In Figure \ref{Fig:redshift_magnitude}, we show the distribution of $W$1 magnitude against SDSS spectroscopic redshift \citep[see][]{2017ApJS..229...39R} for the parent NLSy1 galaxy sample (red) and 520 variable candidates (blue). About 87\% of the variable candidates are located at $z<0.4$ and all of them are brighter than 16 mag in $W$1-band.

For quantitative variability analysis of our final sample of 520 NLSy1 galaxies, we combined data from `AllWISE MultiEpoch Photometry (MEP) Database'' and NEOWISE-R allowing us to study about 7 years long light curves.
Since $W$3 and $W$4 bands data are only available between 2010-2011 in one or two windows, and many photometric points are marked as ``null'', which means no useful brightness estimate could be made, we, therefore, focus on data acquired only in $W$1 and $W$2 bands.
To exclude bad photometric measurements available in the  {\it WISE}
database in $W$1 and $W$2 bands for these 520 objects we adopted the following
criteria

\begin{enumerate}
\item The reduced $\chi^2$ of single-exposure profile-fit is less than $5$ in 
both $W$1 and $W$2 bands  (i.e., \texttt{w1rchi2}$<5$ and \texttt{w2rchi2}$<5$), 
\item The  number of point spread function (PSF) components used in profile 
fit for a source (\texttt{nb}) must be less than 3, 
\item The  best quality single-exposure image 
(\texttt{qi\textunderscore fact}=1), and frames are unaffected by known artifacts 
(\texttt{cc\textunderscore flags}=`0000') and are not actively 
deblended (na$=$0),
\item The number of photometric measurements available in a day must be 
at least five. This condition was applied only for variability analysis within a day.
\end{enumerate}

Using the above constraints, from our sample of 520 sources,  we arrived at a 
final sample of 492 NLSy1 galaxies for further analysis. To study intra-day 
variability, we created the intra-day 
light curves by separating the photometric points  into groups 
(hereafter ``days''). One intra-day lightcurve includes all photometric
data points having time gaps between two successive points lesser 
than 1.2 days. It is possible that the WISE images are affected by cosmic rays which are easily identifiable as outliers in the photometric light curves of an object. Therefore,  we applied a 3$\sigma$ (where $\sigma$ is the standard deviation of the light curve) clipping to each intra-day 
light curve to remove photometric outliers usually present in the survey 
data. We found that majority of the light curves do not have any outliers while in some cases they are present but such number of outliers are usually one in a light curve which were removed by our 3$\sigma$ clipping. Thus, the light curves used in the analysis were devoid of the effects of cosmic ray hits. We visually inspected those light curves before removing the outliers. We also noticed that in many light curves photometric points were present with a separation of only about 11s in successive WISE epochs. Since the duration of the \textit{WISE} orbit is about 1.5 hours, we took the mean of the photometric points available within 1.5 hours and constructed the light curve. Sample light curves for six sources in $W$1 and $W$2 bands are shown in 
Figure \ref{Fig:lc} with each box representing a duration of 1.2 days. The light curve of a nearby star present in the same field is over plotted (after shifting by a constant value for the purpose of comparison) along with the light curve of NLSy1 galaxies. The variation of stars are constant over short and long-term while NLSy1 galaxies show large amplitude of variability. The number of windows in a light curve varies from 2 to 14 with a median of 10, while the total number of points range from 9 to 471 with a median of 108.

For all objects, if intra-day light curves have more than five photometric data points, we considered them for short-term variability analysis. The intra-day light curves used in this analysis thus consisted of a minimum of 5 and maximum of about 86 photometric points while long-term light curves are very sparsely sampled. In some of the windows, the number of photometric points can be as low as one because of the other photometric points if any not satisfying the criteria mentioned above. Though such windows were not considered for intra-day variability analysis, they were considered for long-term variability analysis. Such sparsely sampled light curves prevented us to perform any time series analysis, such as auto-correlation analysis to estimate the correlation timescale or cross-correlation analysis between $W$1 and $W$2 bands to estimate the relative size of emitting regions. Thus, in this work, we focused only on the amplitude of variability in $W$1 and $W$2 bands.

\section{Analysis}\label{sec:results}

\subsection{Flux variability}\label{sec:analysis}
To quantitatively study variability, we calculated the intrinsic 
amplitude of variability ($\sigma_m$), which is the variance of the observed light curve after removing the measurement uncertainty \citep[see also][]{2017ApJ...842...96R}. The $\sigma_m$ is calculated using the following formalism described in \citet{2007AJ....134.2236S}.
\begin{equation}
\Sigma=\sqrt{\frac{1}{n-1}\sum_{i=1}^{N}(m_i - <m>)^2},
\end{equation}
where $m_i$ is the magnitude at $i$-th point and $<m>$ is the weighted average.
The amplitude of variability  can be written as  
$\sigma_m$ is             
\[\sigma_m  =
  \begin{cases}
    \sqrt{\Sigma^2 - \epsilon^2},  & \quad \text{if } \Sigma>\epsilon,\\
     0,                            & \quad  \text{otherwise.}\\
  \end{cases}
\]	               
where the error $\epsilon$ is calculated from the individual errors as follows
\begin{equation}
\epsilon^2=\frac{1}{N}\sum_{i=i}^{N}{\epsilon_{i}^{2} + \epsilon_{s}^2}. 
\end{equation}
Here $\epsilon_{i}$ is the measurement uncertainty of $i$-th point and $\epsilon_{s}$ is the systematic uncertainty. \citet{2011ApJ...735..112J} reported systematic uncertainties of 0.024 mag and 0.028 mag in $W$1 and $W$2 respectively. Therefore, this error has been added in quadrature to the measurement uncertainty and thus taking care of systematic as well as random errors. Spurious correlation may occur if $\sigma_m$ is not corrected for the redshift of the object, especially important for a flux-limited sample, we, therefore, calculated rest frame $\sigma_m$ by multiplying $\sigma_m$ with $\sqrt{(1+z)}$, which is based on the power spectral
density of variability having slope 2 \citep[see][]{2009ApJ...698..895K}. However, the majority ($\sim 87\%$) of the variable candidates in our sample is below $z=0.4$, thus, redshift correction is insignificant.

\subsubsection{Long-term variability amplitude}
\begin{figure}
\centering
\resizebox{9cm}{4.2cm}{\includegraphics{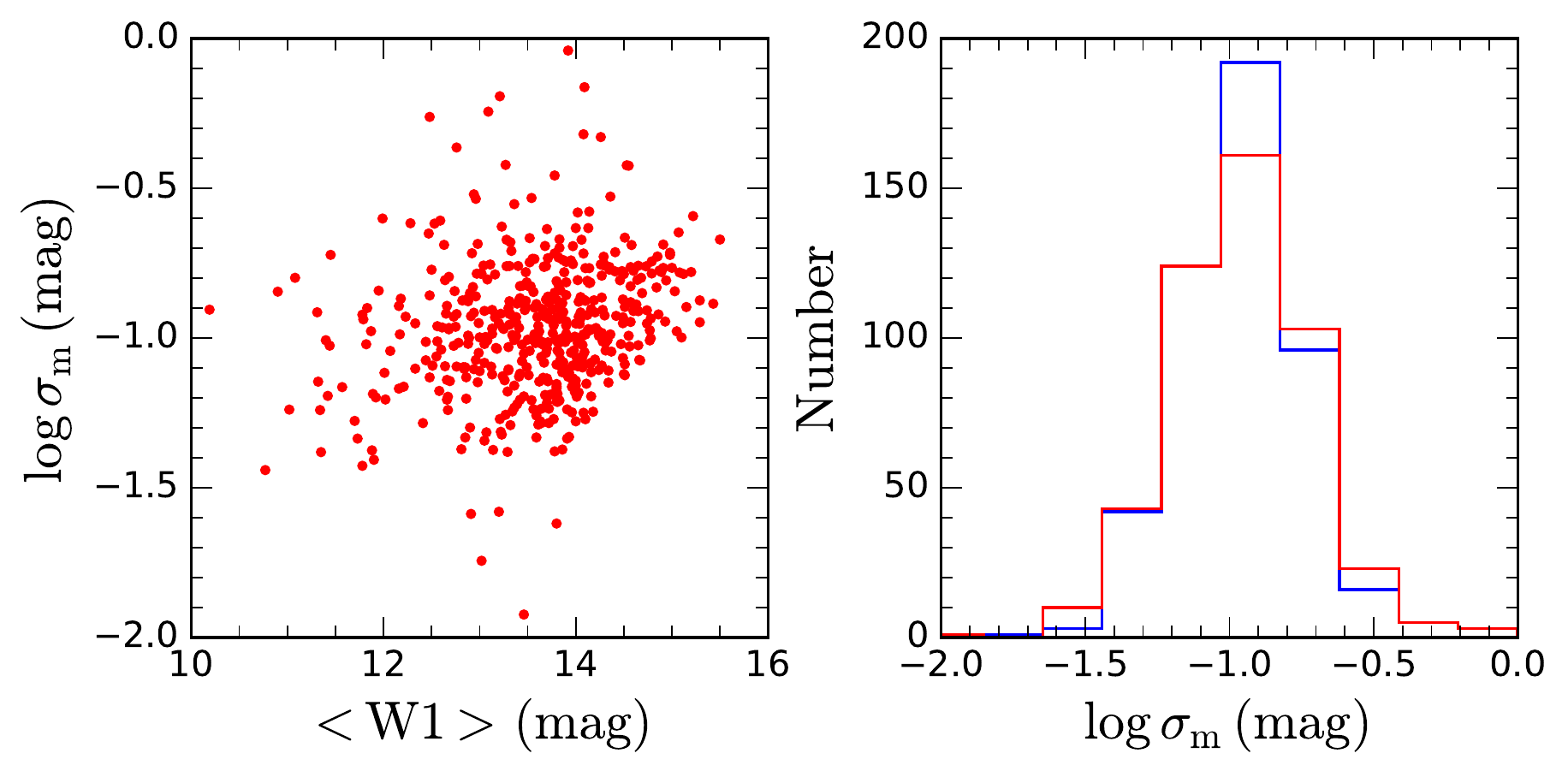}}
\caption{Left: Variability amplitude in $W$1 band in the long-term against mean $W$1 magnitude. Right: The distributions of variability amplitude in $W$1 (blue) and $W$2 (red) bands.}\label{Fig:var}. 
\end{figure}

Variability information of all the 492 candidate variables are provided in Table \ref{Table:var_info}. We found that in the long-term 483 and 473 NLSy1 galaxies have $\sigma_m>0$ in the $W$1 and $W$2 bands respectively. Among these, 473 NLSy1 galaxies\footnote{Light curves of all the 473 NLSy1 galaxies along with the values of their long-term variability amplitude in $W$1 and $W$2 bands can be obtained in the electronic version of the paper.} have $\sigma_m>0$ both in $W$1 and $W$2 bands. The $\sigma_m$ in $W$1 band is plotted against $W$1 magnitude in Figure \ref{Fig:var} on the left panel whereas the distributions of $\sigma_m$ in $W$1 and $W$2 bands are shown in the right panel. The average $\sigma_m$ is $0.11\pm0.07$ mag ($0.11\pm0.08$ mag) in $W$1 ($W$2) band. Thus, there seems to be no difference in the amplitude of flux variations in $W$1 and $W$2. This has also been confirmed by a two-sample K–S test. \citet{2017ApJ...842...96R} estimated long-term variability in $V$-band for a large sample of NLSy1 and BLSy1 galaxies. Cross-matching our sample with them, we found that 142 NLSy1 galaxies are common to this work and \citet{2017ApJ...842...96R}. The average long-term optical variability of these 142 NLSy1 galaxies is 0.11 mag similar to that of their average long-term $W$1 band variability, which is 0.11 mag.

 \begin{table*}
 \caption{Variability information. The columns are as follows: (1) SDSS ID (plate-mjd-fiber), (2) RA (degree), (3) DEC (degree), (4) redshift, (5) log $\lambda L_{5100}$ (erg s$^{-1}$), (6) number of points in $W$1 band light curve, (7) number of points in $W$2 band light curve, (8) number of short-term light curves with minimum 5 points in both $W$1 and $W$2 bands, (9) average $W$1-band magnitude, (10) average $W$2-band magnitude, (11) variability amplitude in $W$1-band (mag), (12) variability amplitude in $W$2-band (mag), and (13) radio power (erg s$^{-1}$). This table is available in its entirety in machine-readable form. A portion is shown here for guidance.}
	\begin{center}
 	\resizebox{0.9\linewidth}{!}{%
     \begin{tabular}{ l l l l l l l l l l l l l}\hline \hline 
    SDSS ID & RA & DEC & z & $\log \lambda L_{\mathrm{5100}}$ & $N_p$ ($W$1) & $N_p$ ($W$2)  &  $N_{lc}$  & $<W1>$ & $<W2>$ & $\sigma_m (W1)$ & $\sigma_m (W2)$ & $\log \, P_{1.4GHz}$\\ 
    (1)            & (2)      & (3)     & (4)    & (5)   & (6) & (7) & (8) & (9)   & (10) & (11) & (12) & (13)\\ \hline
   2004-53737-0466 & 182.4384 & 32.2836 & 0.1447 & 44.04 & 96 & 97 & 9 & 12.64 & 11.63 & 0.16 & 0.13 & 39.18\\
   0873-52674-0158 & 155.5400 & 48.3538 & 0.0626 & 42.85 & 54 & 54 & 3 & 13.70 & 13.29 & 0.09 & 0.13 & ... \\
   0443-51873-0092 & 127.7197 & 48.3332 & 0.2233 & 43.19 & 107 & 107 & 9 & 14.14 & 13.18 & 0.26 & 0.22	& ... \\
   
     \hline \hline
        \end{tabular} } 
        \label{Table:var_info}
        \end{center}
    \end{table*}

\subsubsection{Short-term variability amplitude}
Figure \ref{Fig:lc} shows examples of light curves in which the short-term flux variations in some objects at some windows are visible. Firstly, the amplitude of short-term variations are quite low compared to the long-term
variations and secondly, in many cases, there is no one to one correspondence (both
in the pattern and amplitude) between the short-term variations in $W$1 and $W$2 bands. 
Such lack of one to one correspondence precludes{\tiny } us to use all the intra-day light curves
for short-term variability analysis. Therefore, to quantify short-term variability we calculated the variability covariance \citep[see][]{2016ApJ...817..119K,2018MNRAS.476.1111P} as
\begin{equation}
C_{12}=\frac{1}{N-1} \sum_{i=1}^{N}(m[W1]_i - <m[W1]>)(m[W2]_i - <m[W2]>)
\end{equation}
and Pearson's correlation coefficient of variability between $W$1 and $W$2 bands as
\begin{equation}
r_{12}=\frac{C_{12}}{\Sigma_{W1} \Sigma_{W2}}
\end{equation}
The value of $r_{12}$ ranges from $-1$ (perfect anti-correlation) to 1 (perfect correlation). A truly variable light curve should have high correlation. Thus, we considered a short-term light curve is variable if $\sigma_m>0$ and $r_{12}>0.8$ \citep{2016ApJ...817..119K,2018MNRAS.476.1111P}. For a given object, we then  calculated average variability from the light curves having $\sigma_m>0$ and $r_{12}>0.8$ as a representation of the intra-day variability for that object. Only 31 and 24 objects showed short-term variability in at least one window in $W$1 and $W$2 bands respectively while only 17 objects showed short-term variability in both $W$1 and $W$2 bands.

We also estimated the duty cycle \citep[DC; see][]{2017ApJ...835..275R} i.e. the fraction of time when an object is variable in intra-day timescale following \citet{1999A&AS..135..477R},
\begin{equation}
DC = 100 \frac{\sum_{i=1}^{n} N_i (1/\Delta t_i)}{\sum_{i=1}^{n} (1/\Delta t_i)} \%,
\end{equation}     
where $\Delta t_i=\Delta t_{i, \mathrm{obs}} (1+z)^{-1}$ is the rest frame duration of the monitoring session of the source on the $i$-th light curve and $n$ is the maximum number of short-term light curves in a given object. If $\sigma_m>0$ and $r_{12}>0.8$ in a short-term light curve then $N_i=1$ else $N_i=0$. The DC for those 17 objects having short-term variation in both the $W$1 and $W$2 bands ranges from 6\% to 67\% with a median of 11\%. The highest DC of 67\% is observed in 3FGL J0948.8+0021, which is a $\gamma$-ray emitting NLSy1 galaxy.

\subsection{Color variation in long-term}\label{sec:spec_index}

\begin{figure}
\centering
\resizebox{9cm}{3.2cm}{\includegraphics{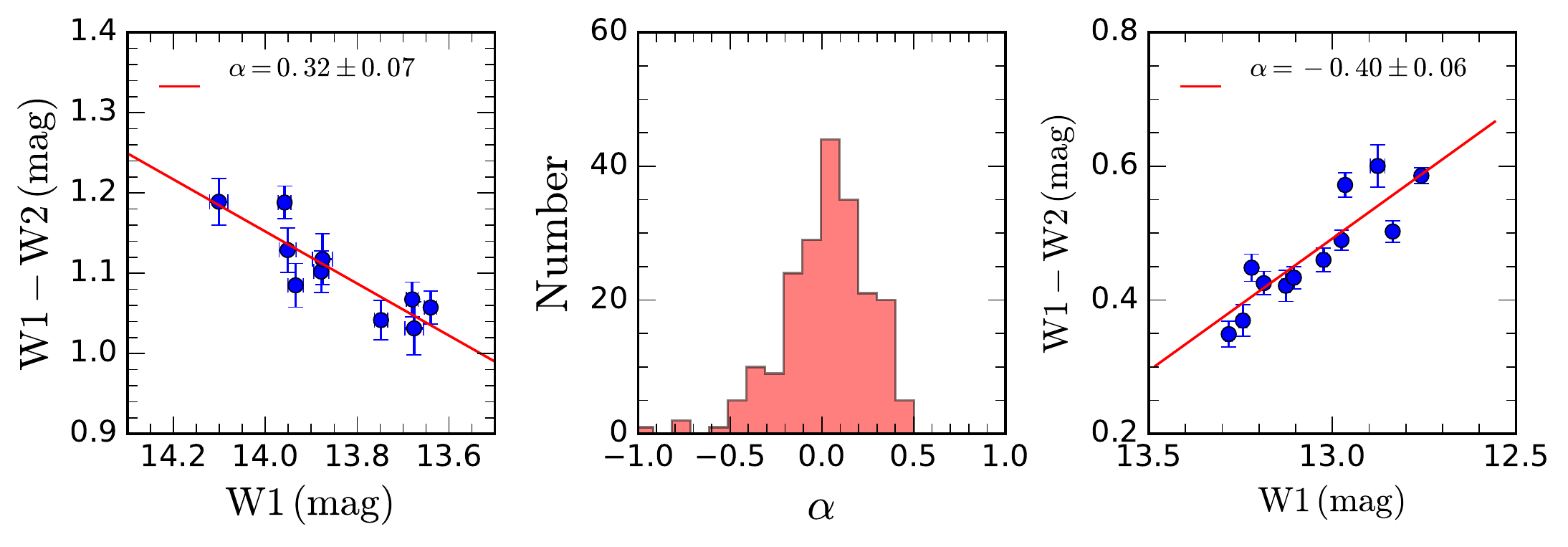}}
\caption{Long-term color variation: an example of bluer when brighter trend (left), the distribution of slope of the objects in our sample (middle) and an example of redder when brighter trend (right).}\label{Fig:index_long}. 
\end{figure}

To investigate color variation in long-term, we calculated the mean color ($<W1-W2>$) and the mean $W$1 magnitude ($<W1>$) by taking the mean of all the data points in each short-term light curves. An example of the fitting to the long-term color-magnitude relation for a NLSy1 galaxy is shown in Figure \ref{Fig:index_long} (left panel), which shows bluer when brighter (BWB) trend. We performed linear fits to the color-magnitude relation taking into account the errors on both axes and calculated the Pearson rank correlation coefficient ($r_p$) between color-magnitude relation of the 232 sources having $\sigma_m>0.1$ mag in the long-term. We plot the distribution of the slope ($\alpha$) of 206 sources, which has $|r_p|>0.8$, in the middle panel. The distribution of $\alpha$ has a mean of 0.05 with a dispersion of 0.24. We note a mean uncertainty of 0.1 in the measurement of $\alpha$. Thus, the dispersion among true source $\alpha$ values is $\sim 0.22$.

About 31\% sources show no color variation within the 1$\sigma$ measurement uncertainty. Among the rest of the 69\% sources, 42\% show BWB trend while 27\% exhibit redder when brighter (RWB) trend. One example of RWB trend seen in a NLSy1 galaxy is shown in the right panel of Figure \ref{Fig:index_long}. The observed IR emission is predominantly due to reprocessed UV/optical emission by the torus. The total observed UV/optical flux is a sum of a constant component that comes from stars in the host galaxy and a variable component from the central nuclear region of AGN. For AGN, with no dilution due to star light from host galaxies, $W1-W2$ 
colors are redder upto $z \sim 3.5$, however, an increased contribution of 
star light will lead to bluer $W1 - W2$ colors \citep{2012ApJ...753...30S}. When the variable nuclear component is dominant, the constant host galaxy component will have little effect on the total observed flux. The increased brightness of the AGN can thus lead to a strong contribution by the AGN dust torus to the observed IR emission leading to a redder $W1-W2$ color when the AGN is brighter. Alternatively, the host galaxy light can play a role to the observed flux only when the AGN fades and the AGN central nuclear luminosity becomes comparable to the host galaxy stellar brightness, which could lead to a BWB trend \citep{2018ApJ...862..109Y}.

\begin{figure*}
\centering
\resizebox{16cm}{6.5cm}{\includegraphics{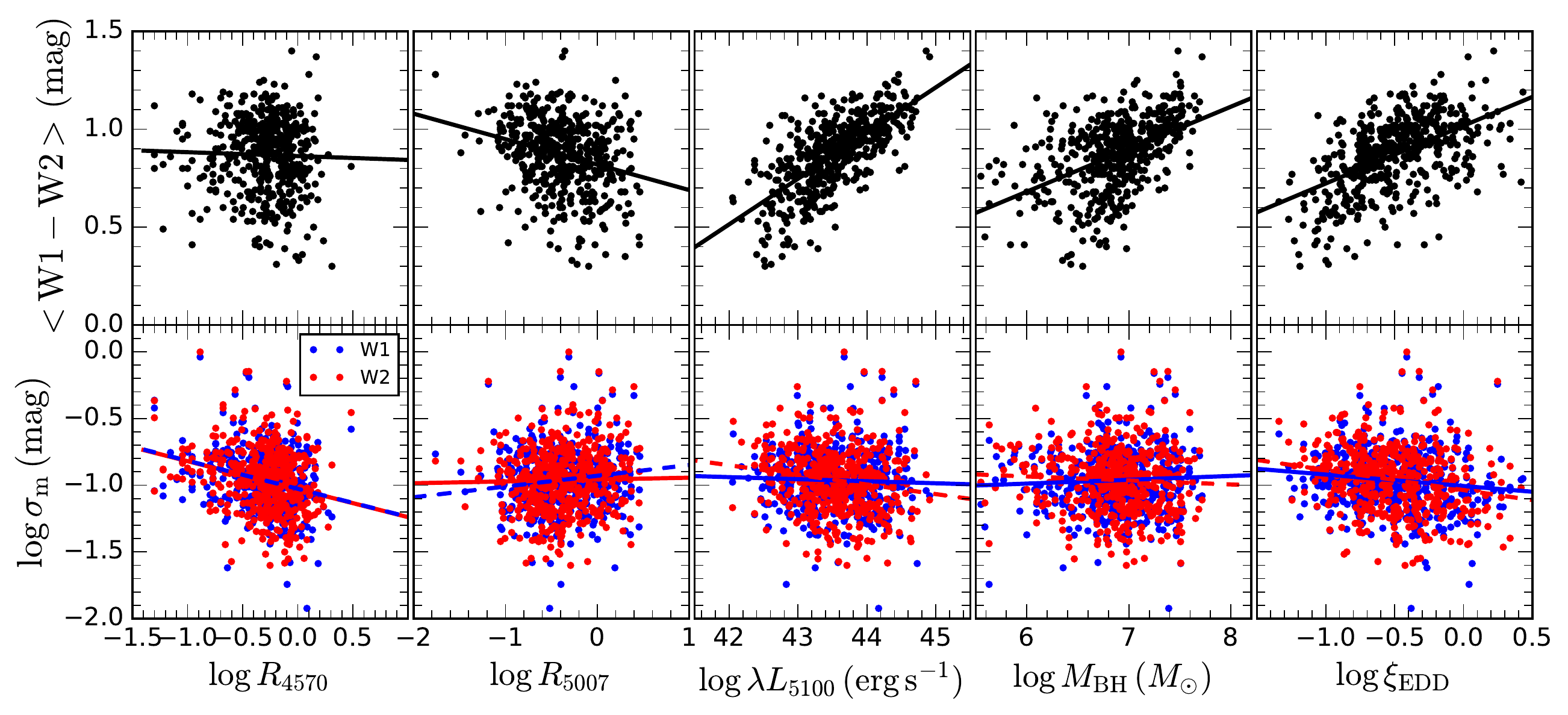}}
\caption{Correlation of average long-term color $<W1-W2>$ (upper panels) and long-term $\sigma_\mathrm{m}$ (lower panels) with $R_{4570}$, $R_{5007}$, $\lambda L_{\mathrm{5100}}$, $M_{\mathrm{BH}}$, $\xi_{\mathrm{Edd}}$ are plotted from left to right columns respectively. The straight lines represent linear fits to the relations. In the lower panel, $\sigma_\mathrm{m}$ for $W$1 (blue) and $W$2 bands (red) are plotted.}\label{Fig:correlation}. 
\end{figure*}

 \begin{table*}
 \caption{Correlation of average color and long-term $\sigma_m$ with different AGN parameters. The columns are as follows: (1) test parameter. Columns (2)$-$(6) note the Spearman correlation coefficient $r_s$ (the $p$-value of no correlation) for $\log R_{4570}$, $\log R_{5007}$, $\log \lambda L_{\mathrm{5100}}$, $\log M_{\mathrm{BH}}/M_{\odot}$ and $\log \xi_{\mathrm{Edd}}$.}
	\begin{center}
 	\resizebox{0.9\linewidth}{!}{%
     \begin{tabular}{ l l l l l l}\hline \hline 
      
    Test parameter & $\log R_{4570}$ & $\log R_{5007}$ & $\log \lambda L_{\mathrm{5100}}$ & $\log M_{\mathrm{BH}}/M_{\odot}$  &  $\log \xi_{\mathrm{Edd}}$ \\ 
    (1)  & (2) & (3) & (4) & (5) & (6) \\ \hline
    $<W1 - W2>$              & $-$0.05 (2e$-$01)  & $-$0.28 (3e$-$10)  & $+$0.69 (2e$-$72) & $+$0.51 (6e$-$35)  & $+$0.54 (3e$-$38)  \\
    $\log \sigma_m$ ($W$1) & $-$0.28 (7e$-$10)  & $+$0.03 (5e$-$01)  & $-$0.06 (1e$-$01) & $+$0.00 (5e$-$01)  & $-$0.14 (1e$-$03)   \\
    $\log \sigma_m$ ($W$2) & $-$0.24 (1e$-$07)  & $+$0.13 (4e$-$03)  & $-$0.18 (3e$-$05) & $-$0.08 (7e$-$02)  & $-$0.21 (2e$-$06) \\ 
     \hline \hline
        \end{tabular} } 
        \label{Table:corr}
        \end{center}
    \end{table*}

 \subsection{Dependence of color and long-term variability on AGN parameters}\label{sec:correlation}
 
 To study the correlation of long-term variability amplitude with AGN parameters, we calculated the black hole mass ($M_{\mathrm{BH}}$), bolometric luminosity ($L_{\mathrm{BOL}}$) and Eddington luminosity ($L_{\mathrm{Edd}}$) based on the spectral fitting results of \citet{2017ApJS..229...39R} following \citet{2017ApJ...842...96R}. First, the size of the broad line region ($R_{\mathrm{BLR}}$) was calculated based on the size-luminosity scaling relation obtained by \cite{2013ApJ...767..149B}. Then assuming virial relationship and spherical distribution of clouds (virial scale factor, $f = 3/4$), the black hole mass of each source was calculated using the full 
width at half maximum of the broad component of H$\beta$ emission line 
($\Delta v$), $M_{\mathrm{BH}} = f R_{\mathrm{BLR}} \Delta v^2/G$. 
The $L_{\mathrm{BOL}}$ was approximated as $L_{\mathrm{BOL}}= 9 \times \lambda L_{\lambda} \mathrm{(5100)} \, \mathrm{erg\, s^{-1}}$ and $L_{\mathrm{Edd}}$ was calculated as $L_{\mathrm{Edd}} = 1.3 \times 10^{38}\, M_{\mathrm{BH}}/M_{\odot} \, \mathrm{erg\, s^{-1}}$ \citep{2000ApJ...533..631K}. 
The Eddington ratio is given by $\xi_{\mathrm{Edd}} = L_{\mathrm{BOL}}/L_{\mathrm{Edd}}$. The values of Fe II 
strength ($R_{4570}$) i.e., the flux ratio of Fe II to H$\beta$ and $R_{5007}$ 
(defined by the flux ratio of [O III] to {total} H$\beta$) were taken from \citet{2017ApJS..229...39R}. 
  
In Figure \ref{Fig:correlation}, we performed several correlations to understand the dependency of mean long-term $W1-W2$ color (upper panels) and long-term variability
amplitude $\sigma_m$ (lower panels) with various physical parameters of the NLSy1 galaxies
in our sample such as $R_{4570}$, $R_{5007}$,  $\lambda L_{\mathrm{5100}}$, 
$M_{\mathrm{BH}}$ and $\xi_{\mathrm{Edd}}$. Linear least square fits (indicated by straight lines) carried out on the data are shown in 
Figure \ref{Fig:correlation} and the results of the Spearman rank correlation analysis are 
summarized in Table \ref{Table:corr} where correlation coefficient ($r_s$) and 
the $p$-value of no correlation is given. We found no correlation between
 $W1-W2$ and $R_{4570}$, however, $W1-W2$ color is inversely 
correlated with $R_{5007}$ having $r_s=-0.28$. 
The $\lambda L_{\mathrm{5100}}$, $M_{\mathrm{BH}}$ and $\xi_{\mathrm{Edd}}$ are strongly correlated with $W1-W2$ color having $r_s=+0.69$, $+0.51$ and $+0.54$ 
respectively.

Figure \ref{Fig:correlation} (top panel) shows a strong trend for more luminous objects to be redder. Firstly, the continuum luminosity at 5100\AA \, was from the optical spectra (stellar contribution subtracted) of the sources and in that epoch, the object can be of any brightness level (faint, moderate or bright). Secondly, the $<W1 - W2>$ color is the average $W1- W2$ of all the points in a particular object.  Thus the middle panels of Figure \ref{Fig:correlation} depict the average behavior of  the NLSy1 galaxy population in $W1-W2$ versus 5100\AA \, luminosity. Thus on average brighter NLSy1 galaxies have redder $W1-W2$ color. Such a behavior is caused by a larger contribution of the dust torus to the observed IR emission at bright flux levels of NLSy1 galaxies \citep{2018ApJ...862..109Y}.

We also investigated the correlation between the $W1-W2$ color against various emission line parameters of the sources. We found an inverse correlation between $W1-W2$ versus $R_{5007}$, which agrees with the findings of \citet{2017NewA...54...30C}. Between $W1-W2$ and $R_{4570}$ we found no correlation, however, \citet{2017NewA...54...30C} noticed a weak positive
correlation between $W1-W2$ and $R_{4570}$. From \citet{2017ApJS..229...39R} using their complete sample of NLSy1 galaxies it is known that both the luminosity of H$\beta$ and [O III] lines correlate positively with the continuum luminosity at 5100\AA \, as
\begin{equation}
\log L(\mathrm{H}\beta_{tot})=-(4.30 \pm 0.21) + (1.056 \pm 0.004) \log \lambda L_{5100}
\label{eq:hb}
\end{equation}
where $L(\mathrm{H}\beta_{tot})$ is the total H$\beta$ luminosity, and  
\begin{equation}
\log L\mathrm{[O III]} = (9.87 \pm 0.29) + (0.721 \pm 0.006) \log \lambda L_{5100}. 
\label{eq:oiii}
\end{equation}
Rearranging equations \ref{eq:hb} and \ref{eq:oiii} we found
\begin{equation}
\log R_{5007} = (14.17 \pm 0.35) - (0.335 \pm 0.007) \log \lambda L_{5100}
\end{equation}
Thus, on average NLSy1 galaxies at higher optical luminosities have lower $R_{5007}$ and redder $W1-W2$. The different relation of $W1-W2$ color versus $R_{4570}$ and $W1-W2$ color versus $R_{5007}$ is not surprising considering their origin. The [O III] emission comes from the narrow line region, while Fe II and H$\beta$ originate in the BLR.

The correlation of $\sigma_m$ in the long-term light curve of $W$1 (blue) and 
$W$2 (red) bands is plotted with $R_{4570}$, $R_{5007}$,  $\lambda L_{\mathrm{5100}}$, 
$M_{\mathrm{BH}}$ and $\xi_{\mathrm{Edd}}$ in the lower panel of Figure 
\ref{Fig:correlation}. There is a weak negative correlation between 
$\sigma_m$ in long-term $W$1 and $W$2 light curves with $R_{4570}$, 
$\lambda L_{\mathrm{5100}}$ and $\xi_{\mathrm{Edd}}$. The results of the Spearman rank 
correlation analysis are given in Table \ref{Table:corr}. This plot can be 
compared with the Figure 8 of \citet{2017ApJ...842...96R}, where the correlation
of $\sigma_m$ in optical has been plotted with the same parameters. 
Interestingly, the correlations which are present in the optical are also 
present here though weaker. This could be simply due to the different origin 
of near-IR variability than optical. However, note that {\it WISE} light curves are very sparsely sampled thus local monthly variation cannot be probed.  

The optical continuum originates from the accretion disk while near-IR 
continuum is mainly due to the radiation from a dusty torus beyond the dust 
sublimation radius \citep{1993ApJ...402..441L}. The optical 
and IR emissions are causally connected and it is not surprising that the 
correlations between variability and various physical characteristics of 
the sources seen in the optical bands are also seen in the IR bands.
The $\xi_{\mathrm{Edd}}$, which is considered to be the main driver of optical 
variability inversely correlates with the $\sigma_m$ in
optical \citep[see][]{2009ApJ...698..895K,2010ApJ...721.1014M,2011A&A...525A..37M,2017ApJ...842...96R}. Such an inverse correlation could be explained using simple standard accretion disk model \citep{1973A&A....24..337S}. The radius ($R$ in the unit of the Schwarzschild radius) of the emission region increases with Eddington ratio for a given wavelength, 
$R \sim T^{-4/3} \sim \left(\dot{m}/M_{BH} \right)^{1/3} \lambda^{4/3}$, 
$\lambda$ is the wavelength of observation and $\dot{m}$ is the mass accretion 
rate in the unit of Eddington rate. At any given wavelength, for a high Eddington ratio object, the primary emission comes from the outer region of the accretion disk, compared to a low  Eddington ratio object, where the primary emission comes from the inner region of the accretion disk.  Thus it is natural to expect an anti-correlation between variability amplitude and Eddington ratio. However, as the observed IR emission is the reprocessed primary emission from the accretion disk, the observed
variability is diluted and therefore the anticorrelation between $\sigma_m$ and $\lambda_{\mathrm{Edd}}$ is much weaker in the IR than optical.

\subsection{Variability of radio subsample}\label{sec:radio}

In the population of NLSy1 galaxies, a minority of about 7\% are radio-loud \citep{2006AJ....132..531K,2017ApJS..229...39R}. To understand the influence of radio emission on the observed color and IR flux variations, we performed a comparative analysis of the color and flux variability characteristics of 
radio-quiet and radio-loud NLSy1 galaxies. 
Towards this, we cross-correlated
our sample of NLSy1 galaxies with the FIRST (Faint Images of the Radio
Sky at Twenty cm) catalog \citep{1995ApJ...450..559B}. This leads us to 63 radio-detected NLSy1 galaxies.

\begin{figure}
\centering
\resizebox{9cm}{4.5cm}{\includegraphics{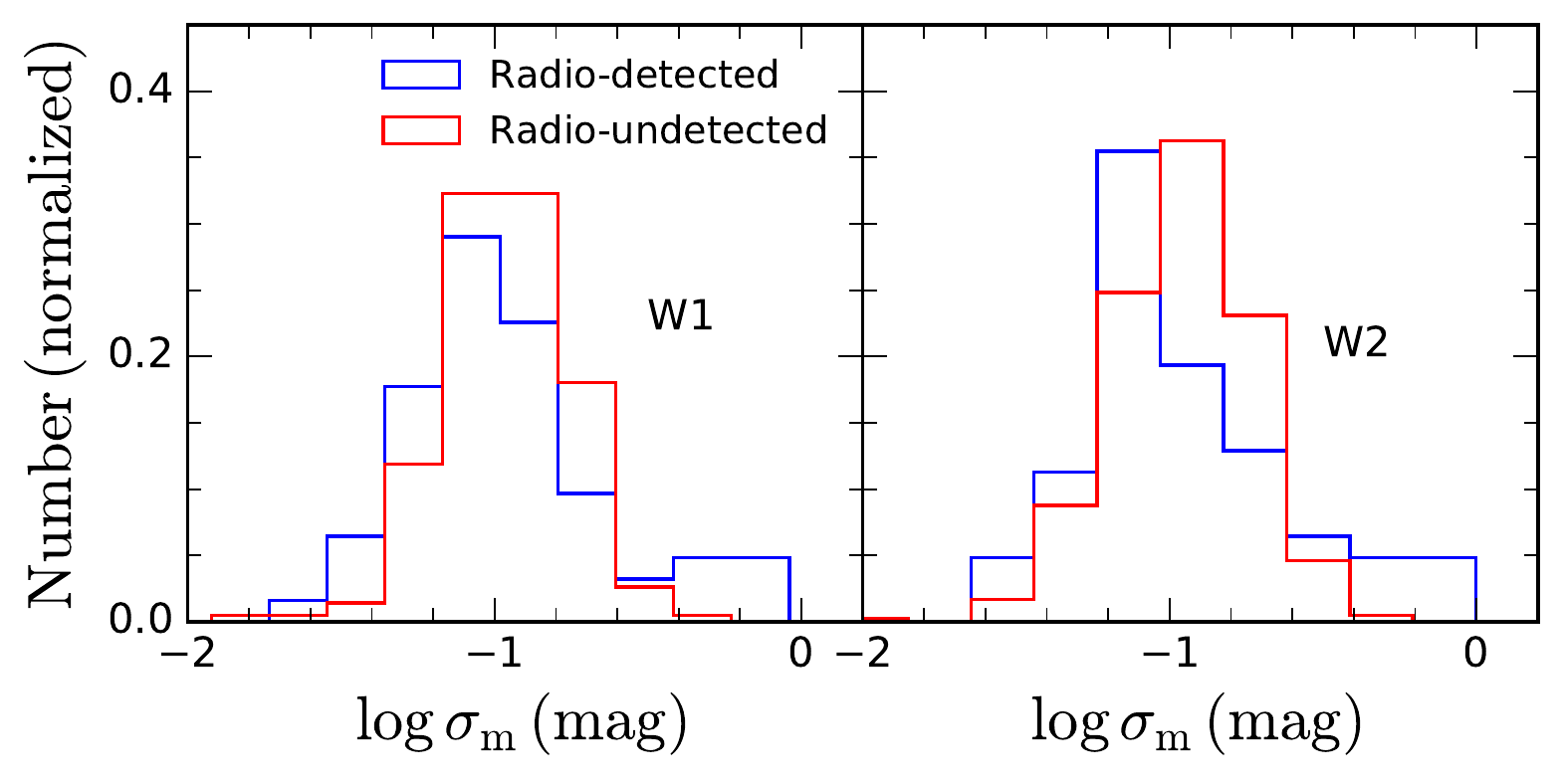}}
\caption{Distribution of long-term variability amplitude of radio-detected (blue) and radio-undetected (red) NLSy1 galaxies.}\label{Fig:radio_var}. 
\end{figure}

\begin{figure}
\centering
\resizebox{9cm}{4.5cm}{\includegraphics{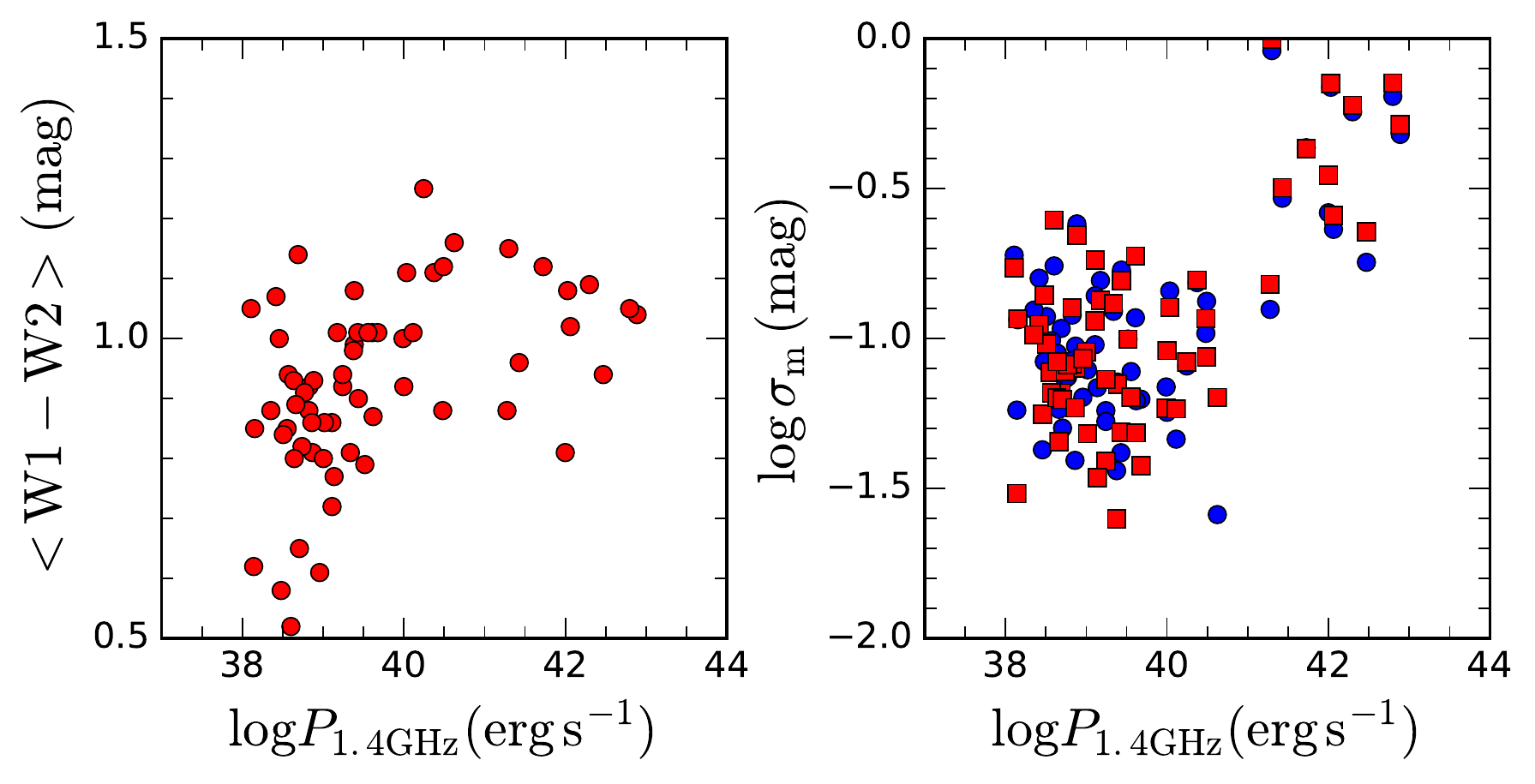}}
\caption{Correlation between mean $W1-W2$ color versus $P_{1.4GHz}$ (left) and $\sigma_m$ versus $P_{\mathrm{1.4GHz}}$ in long-term (right). The circles and squares in the right panel represent $W$1 and $W$2 bands respectively.}\label{Fig:radio_corr}. 
\end{figure}

The distribution of $\sigma_m$ for radio-detected and radio-undetected NLSy1 galaxies are plotted in Figure \ref{Fig:radio_var}. Considering long-term variability, the mean $\sigma_m$ is $0.16 \pm 0.16$ mag and $0.12 \pm 0.06$ mag for the radio-detected and undetected sample respectively both in $W$1 and $W$2 bands. A two-sample
KS test also confirms a marginal difference in the IR variability
between radio-detected and radio-undetected NLSy1 galaxies having a $D-$ statistics value 0.15 ($p$-value=0.17) and 0.18 ($p$-value=0.06) in the long-term in $W$1 and $W$2 bands. We plot in Figure \ref{Fig:radio_corr} (left panel) the mean $W1-W2$ color of 
radio detected NLSy1 galaxies against their radio power at 1.4 GHz 
($P_{1.4\mathrm{GHz}}$). A positive correlation is found with a Spearman correlation 
coefficient  $r_s=0.5$ and $p$-value=$3e-5$ respectively.  A positive 
correlation (right panel) is also found between the long-term variability 
amplitude and radio power at 1.4 GHz with $r_s=0.33$ ($r_s=0.40$) and 
$p-$ value=$7e-03 \,(1e-03)$ for $W$1 ($W$2) bands respectively. Note that a similar correlation between long-term variability amplitude and radio power has also been observed in the optical band \citep{2017ApJ...842...96R}. Interestingly, 9 objects with radio power $P_{\mathrm{1.4GHz}}>10^{41} \mathrm{erg \,s^{-1}}$ exhibit stronger variability with $\sigma_{\mathrm{m}} >0.2$ mag (also see Figure \ref{Fig:radio_var}). This includes 8 $\gamma$-ray emitting NLSy1 galaxies detected by {\it Fermi}-Large Area Telescope. Thus, though the major component of the IR variation in NLSy1 galaxies is the dusty torus, in the case of radio-detected NLSy1 galaxies, non-thermal jet emission too contributes to the observed IR emission as found from the strong trend of increasing variability amplitude with radio power.

\section{Summary}\label{sec:conclusion}
We have carried out a systematic analysis of the IR color and 
flux variability of a large sample of NLSy1 galaxies in two IR bands, namely $W$1 and $W$2. We find only $\sim 5\%$ of the initial WISE detected sample of 10,777 NLSy1 galaxies show detectable variability. The main findings of the color and variability analysis of the variable NLSy1 galaxies are summarized below 
\begin{itemize}
\item In the {\it WISE} color-color diagram, 57.6\% objects of the 520 candidate variable NLSy1 galaxies fall within the WISE Gamma-ray strips having a mean $W1-W2$ color of $0.99\pm 0.18$ mag.

\item In long-term, the mean amplitude of variability in $W$1 and $W$2 bands
is $0.11\pm 0.07$ mag and $0.11\pm 0.08$ mag respectively. The amplitude of variability on long (year like) timescales is larger than that obtained on short timescales. The long-term IR variability 
characteristics of NLSy1 galaxies is similar to their long-term optical
variability characteristics \citep{2017ApJ...842...96R}. This is expected as the observed IR emission is the re-processed optical/UV emission
from the central engine of NLSy1 galaxies.
\item The average $W1-W2$ color is anti-correlated with $R_{5007}$ but 
strongly correlated with $\lambda L_{\mathrm{5100}}$, $M_{\mathrm{BH}}$ and 
$\xi_{\mathrm{Edd}}$. However, no correlation was found between $W1-W2$ color 
and $R_{4570}$.
\item Weak negative correlation was observed between $\sigma_m$ with 
$R_{4570}$, $\lambda L_{\mathrm{5100}}$, and $\xi_{\mathrm{Edd}}$ in long-term. This is similar
to what has been noticed in the optical band though here the correlations are much weaker compared to the optical. Densely sampled long-term light curves could provide more insights into the underlying physics of IR flux variation. 
\item The radio-detected NLSy1 galaxies on average showed similar variability compared to the radio-undetected sample. Considering only the radio-detected NLSy1 galaxies, a positive 
correlation is found between the mean $W1-W2$ color and the radio power 
at 1.4 GHz. A total of 9 objects with radio power $P_{1.4GHz} > 10^{41}$ erg s$^{-1}$, which includes 8 $\gamma$-ray emitting NLSy1 galaxies exhibit stronger variability ($\sigma_m>0.2$ mag). Also, for the same sample, a positive correlation is noticed
between the amplitude of variability and the 1.4 GHz radio power.  This points to the possible contribution of jet emission
to the observed IR emission in the radio-detected NLSy1 galaxies.
\end{itemize}

\section{Acknowledgments}
We thank the referee  for his valuable comments and suggestions which helped to improve the manuscript. S.R. acknowledges the support by the Basic Science Research Program through the National Research Foundation of Korea government (2016R1A2B3011457). A.J. thanks the Science Academies, for the fellowship provided to her to carry out a summer project at IIA from the IASc-INSA-NASI Summer Research Fellowship $-$ 2017. S.R. thanks Neha Sharma (KHU, South Korea) for carefully reading the manuscript. This publication makes use of data products from the Wide-field Infrared Survey Explorer, which is a joint project of the University of California, Los Angeles, and the Jet Propulsion Laboratory/California Institute of Technology, funded by the National Aeronautics and Space Administration. 


\bibliographystyle{mnras}
\bibliography{ref}

\begin{thebibliography}{}
\makeatletter
\relax
\def\mn@urlcharsother{\let\do\@makeother \do\$\do\&\do\#\do\^\do\_\do\%\do\~}
\def\mn@doi{\begingroup\mn@urlcharsother \@ifnextchar [ {\mn@doi@}
  {\mn@doi@[]}}
\def\mn@doi@[#1]#2{\def\@tempa{#1}\ifx\@tempa\@empty \href
  {http://dx.doi.org/#2} {doi:#2}\else \href {http://dx.doi.org/#2} {#1}\fi
  \endgroup}
\def\mn@eprint#1#2{\mn@eprint@#1:#2::\@nil}
\def\mn@eprint@arXiv#1{\href {http://arxiv.org/abs/#1} {{\tt arXiv:#1}}}
\def\mn@eprint@dblp#1{\href {http://dblp.uni-trier.de/rec/bibtex/#1.xml}
  {dblp:#1}}
\def\mn@eprint@#1:#2:#3:#4\@nil{\def\@tempa {#1}\def\@tempb {#2}\def\@tempc
  {#3}\ifx \@tempc \@empty \let \@tempc \@tempb \let \@tempb \@tempa \fi \ifx
  \@tempb \@empty \def\@tempb {arXiv}\fi \@ifundefined
  {mn@eprint@\@tempb}{\@tempb:\@tempc}{\expandafter \expandafter \csname
  mn@eprint@\@tempb\endcsname \expandafter{\@tempc}}}

\bibitem[\protect\citeauthoryear{{Abdo} et~al.,}{{Abdo}
  et~al.}{2009}]{2009ApJ...699..976A}
{Abdo} A.~A.,  et~al., 2009, \mn@doi [\apj] {10.1088/0004-637X/699/2/976},
  \href {http://cdsads.u-strasbg.fr/abs/2009ApJ...699..976A} {699, 976}

\bibitem[\protect\citeauthoryear{{Alam} et~al.,}{{Alam}
  et~al.}{2015}]{2015ApJS..219...12A}
{Alam} S.,  et~al., 2015, \mn@doi [\apjs] {10.1088/0067-0049/219/1/12}, \href
  {http://adsabs.harvard.edu/abs/2015ApJS..219...12A} {219, 12}

\bibitem[\protect\citeauthoryear{{Antonucci}}{{Antonucci}}{1993}]{1993ARA&A..31..473A}
{Antonucci} R.,  1993, \mn@doi [\araa] {10.1146/annurev.aa.31.090193.002353},
  \href {http://cdsads.u-strasbg.fr/abs/1993ARA%26A..31..473A} {31, 473}

\bibitem[\protect\citeauthoryear{{Baldi}, {Capetti}, {Robinson}, {Laor}  \&
  {Behar}}{{Baldi} et~al.}{2016}]{2016MNRAS.458L..69B}
{Baldi} R.~D.,  {Capetti} A.,  {Robinson} A.,  {Laor} A.,   {Behar} E.,  2016,
  \mn@doi [\mnras] {10.1093/mnrasl/slw019}, \href
  {http://cdsads.u-strasbg.fr/abs/2016MNRAS.458L..69B} {458, L69}

\bibitem[\protect\citeauthoryear{{Becker}, {White}  \& {Helfand}}{{Becker}
  et~al.}{1995}]{1995ApJ...450..559B}
{Becker} R.~H.,  {White} R.~L.,   {Helfand} D.~J.,  1995, \mn@doi [\apj]
  {10.1086/176166}, \href {http://adsabs.harvard.edu/abs/1995ApJ...450..559B}
  {450, 559}

\bibitem[\protect\citeauthoryear{{Bentz} et~al.,}{{Bentz}
  et~al.}{2013}]{2013ApJ...767..149B}
{Bentz} M.~C.,  et~al., 2013, \mn@doi [\apj] {10.1088/0004-637X/767/2/149},
  \href {http://cdsads.u-strasbg.fr/abs/2013ApJ...767..149B} {767, 149}

\bibitem[\protect\citeauthoryear{{Boller}, {Brandt}  \& {Fink}}{{Boller}
  et~al.}{1996}]{1996A&A...305...53B}
{Boller} T.,  {Brandt} W.~N.,   {Fink} H.,  1996, \aap, \href
  {http://adsabs.harvard.edu/abs/1996A%26A...305...53B} {305, 53}

\bibitem[\protect\citeauthoryear{{Calderone}, {Ghisellini}, {Colpi}  \&
  {Dotti}}{{Calderone} et~al.}{2013}]{2013MNRAS.431..210C}
{Calderone} G.,  {Ghisellini} G.,  {Colpi} M.,   {Dotti} M.,  2013, \mn@doi
  [\mnras] {10.1093/mnras/stt157}, \href
  {http://cdsads.u-strasbg.fr/abs/2013MNRAS.431..210C} {431, 210}

\bibitem[\protect\citeauthoryear{{Chen}, {Liu}  \& {Shan}}{{Chen}
  et~al.}{2017}]{2017NewA...54...30C}
{Chen} P.~S.,  {Liu} J.~Y.,   {Shan} H.~G.,  2017, \mn@doi [\na]
  {10.1016/j.newast.2017.01.005}, \href
  {http://adsabs.harvard.edu/abs/2017NewA...54...30C} {54, 30}

\bibitem[\protect\citeauthoryear{{Cutri} et~al.,}{{Cutri}
  et~al.}{2011}]{2011wise.rept....1C}
{Cutri} R.~M.,  et~al., 2011, Technical report, {Explanatory Supplement to the
  WISE Preliminary Data Release Products}

\bibitem[\protect\citeauthoryear{{D'Ammando}, {Orienti}, {Larsson}  \&
  {Giroletti}}{{D'Ammando} et~al.}{2015}]{2015MNRAS.452..520D}
{D'Ammando} F.,  {Orienti} M.,  {Larsson} J.,   {Giroletti} M.,  2015, \mn@doi
  [\mnras] {10.1093/mnras/stv1278}, \href
  {http://adsabs.harvard.edu/abs/2015MNRAS.452..520D} {452, 520}

\bibitem[\protect\citeauthoryear{{Doi}, {Nagai}, {Asada}, {Kameno}, {Wajima}
  \& {Inoue}}{{Doi} et~al.}{2006}]{2006PASJ...58..829D}
{Doi} A.,  {Nagai} H.,  {Asada} K.,  {Kameno} S.,  {Wajima} K.,   {Inoue} M.,
  2006, \mn@doi [\pasj] {10.1093/pasj/58.5.829}, \href
  {http://adsabs.harvard.edu/abs/2006PASJ...58..829D} {58, 829}

\bibitem[\protect\citeauthoryear{{Drake} et~al.,}{{Drake}
  et~al.}{2009}]{2009ApJ...696..870D}
{Drake} A.~J.,  et~al., 2009, \mn@doi [\apj] {10.1088/0004-637X/696/1/870},
  \href {http://cdsads.u-strasbg.fr/abs/2009ApJ...696..870D} {696, 870}

\bibitem[\protect\citeauthoryear{{Giveon}, {Maoz}, {Kaspi}, {Netzer}  \&
  {Smith}}{{Giveon} et~al.}{1999}]{1999MNRAS.306..637G}
{Giveon} U.,  {Maoz} D.,  {Kaspi} S.,  {Netzer} H.,   {Smith} P.~S.,  1999,
  \mn@doi [\mnras] {10.1046/j.1365-8711.1999.02556.x}, \href
  {http://cdsads.u-strasbg.fr/abs/1999MNRAS.306..637G} {306, 637}

\bibitem[\protect\citeauthoryear{{Goodrich}}{{Goodrich}}{1989}]{1989ApJ...342..224G}
{Goodrich} R.~W.,  1989, \mn@doi [\apj] {10.1086/167586}, \href
  {http://adsabs.harvard.edu/abs/1989ApJ...342..224G} {342, 224}

\bibitem[\protect\citeauthoryear{{Grupe}}{{Grupe}}{2004}]{2004AJ....127.1799G}
{Grupe} D.,  2004, \mn@doi [\aj] {10.1086/382516}, \href
  {http://cdsads.u-strasbg.fr/abs/2004AJ....127.1799G} {127, 1799}

\bibitem[\protect\citeauthoryear{{Hoffman}, {Cutri}, {Masci}, {Fowler}, {Marsh}
   \& {Jarrett}}{{Hoffman} et~al.}{2012}]{2012AJ....143..118H}
{Hoffman} D.~I.,  {Cutri} R.~M.,  {Masci} F.~J.,  {Fowler} J.~W.,  {Marsh}
  K.~A.,   {Jarrett} T.~H.,  2012, \mn@doi [\aj] {10.1088/0004-6256/143/5/118},
  \href {http://adsabs.harvard.edu/abs/2012AJ....143..118H} {143, 118}

\bibitem[\protect\citeauthoryear{{Jarrett} et~al.,}{{Jarrett}
  et~al.}{2011}]{2011ApJ...735..112J}
{Jarrett} T.~H.,  et~al., 2011, \mn@doi [\apj] {10.1088/0004-637X/735/2/112},
  \href {http://adsabs.harvard.edu/abs/2011ApJ...735..112J} {735, 112}

\bibitem[\protect\citeauthoryear{{Jiang} et~al.,}{{Jiang}
  et~al.}{2012}]{2012ApJ...759L..31J}
{Jiang} N.,  et~al., 2012, \mn@doi [\apjl] {10.1088/2041-8205/759/2/L31}, \href
  {http://cdsads.u-strasbg.fr/abs/2012ApJ...759L..31J} {759, L31}

\bibitem[\protect\citeauthoryear{{Kaspi}, {Smith}, {Netzer}, {Maoz}, {Jannuzi}
  \& {Giveon}}{{Kaspi} et~al.}{2000}]{2000ApJ...533..631K}
{Kaspi} S.,  {Smith} P.~S.,  {Netzer} H.,  {Maoz} D.,  {Jannuzi} B.~T.,
  {Giveon} U.,  2000, \mn@doi [\apj] {10.1086/308704}, \href
  {http://cdsads.u-strasbg.fr/abs/2000ApJ...533..631K} {533, 631}

\bibitem[\protect\citeauthoryear{{Kelly}, {Bechtold}  \&
  {Siemiginowska}}{{Kelly} et~al.}{2009}]{2009ApJ...698..895K}
{Kelly} B.~C.,  {Bechtold} J.,   {Siemiginowska} A.,  2009, \mn@doi [\apj]
  {10.1088/0004-637X/698/1/895}, \href
  {http://adsabs.harvard.edu/abs/2009ApJ...698..895K} {698, 895}

\bibitem[\protect\citeauthoryear{{Komossa}, {Voges}, {Xu}, {Mathur}, {Adorf},
  {Lemson}, {Duschl}  \& {Grupe}}{{Komossa} et~al.}{2006}]{2006AJ....132..531K}
{Komossa} S.,  {Voges} W.,  {Xu} D.,  {Mathur} S.,  {Adorf} H.-M.,  {Lemson}
  G.,  {Duschl} W.~J.,   {Grupe} D.,  2006, \mn@doi [\aj] {10.1086/505043},
  \href {http://cdsads.u-strasbg.fr/abs/2006AJ....132..531K} {132, 531}

\bibitem[\protect\citeauthoryear{{Koshida} et~al.,}{{Koshida}
  et~al.}{2014}]{2014ApJ...788..159K}
{Koshida} S.,  et~al., 2014, \mn@doi [\apj] {10.1088/0004-637X/788/2/159},
  \href {http://adsabs.harvard.edu/abs/2014ApJ...788..159K} {788, 159}

\bibitem[\protect\citeauthoryear{{Koz{\l}owski}, {Kochanek}, {Ashby}, {Assef},
  {Brodwin}, {Eisenhardt}, {Jannuzi}  \& {Stern}}{{Koz{\l}owski}
  et~al.}{2016}]{2016ApJ...817..119K}
{Koz{\l}owski} S.,  {Kochanek} C.~S.,  {Ashby} M.~L.~N.,  {Assef} R.~J.,
  {Brodwin} M.,  {Eisenhardt} P.~R.,  {Jannuzi} B.~T.,   {Stern} D.,  2016,
  \mn@doi [\apj] {10.3847/0004-637X/817/2/119}, \href
  {http://adsabs.harvard.edu/abs/2016ApJ...817..119K} {817, 119}

\bibitem[\protect\citeauthoryear{{Kshama}, {Paliya}  \& {Stalin}}{{Kshama}
  et~al.}{2017}]{2017MNRAS.466.2679K}
{Kshama} S.~K.,  {Paliya} V.~S.,   {Stalin} C.~S.,  2017, \mn@doi [\mnras]
  {10.1093/mnras/stw3317}, \href
  {http://adsabs.harvard.edu/abs/2017MNRAS.466.2679K} {466, 2679}

\bibitem[\protect\citeauthoryear{{Laor} \& {Draine}}{{Laor} \&
  {Draine}}{1993}]{1993ApJ...402..441L}
{Laor} A.,  {Draine} B.~T.,  1993, \mn@doi [\apj] {10.1086/172149}, \href
  {http://adsabs.harvard.edu/abs/1993ApJ...402..441L} {402, 441}

\bibitem[\protect\citeauthoryear{{Leighly}}{{Leighly}}{1999a}]{1999ApJS..125..297L}
{Leighly} K.~M.,  1999a, \mn@doi [\apjs] {10.1086/313277}, \href
  {http://cdsads.u-strasbg.fr/abs/1999ApJS..125..297L} {125, 297}

\bibitem[\protect\citeauthoryear{{Leighly}}{{Leighly}}{1999b}]{1999ApJS..125..317L}
{Leighly} K.~M.,  1999b, \mn@doi [\apjs] {10.1086/313287}, \href
  {http://cdsads.u-strasbg.fr/abs/1999ApJS..125..317L} {125, 317}

\bibitem[\protect\citeauthoryear{{Li}, {McGreer}, {Wu}, {Fan}  \& {Yang}}{{Li}
  et~al.}{2018}]{2018arXiv180507747L}
{Li} Z.,  {McGreer} I.~D.,  {Wu} X.-B.,  {Fan} X.,   {Yang} Q.,  2018,
  preprint, \href {http://adsabs.harvard.edu/abs/2018arXiv180507747L} {}
  (\mn@eprint {arXiv} {1805.07747})

\bibitem[\protect\citeauthoryear{{Liu}, {Wang}, {Mao}  \& {Wei}}{{Liu}
  et~al.}{2010}]{2010ApJ...715L.113L}
{Liu} H.,  {Wang} J.,  {Mao} Y.,   {Wei} J.,  2010, \mn@doi [\apjl]
  {10.1088/2041-8205/715/2/L113}, \href
  {http://cdsads.u-strasbg.fr/abs/2010ApJ...715L.113L} {715, L113}

\bibitem[\protect\citeauthoryear{{Lynden-Bell}}{{Lynden-Bell}}{1969}]{1969Natur.223..690L}
{Lynden-Bell} D.,  1969, \mn@doi [\nat] {10.1038/223690a0}, \href
  {http://cdsads.u-strasbg.fr/abs/1969Natur.223..690L} {223, 690}

\bibitem[\protect\citeauthoryear{{MacLeod} et~al.,}{{MacLeod}
  et~al.}{2010}]{2010ApJ...721.1014M}
{MacLeod} C.~L.,  et~al., 2010, \mn@doi [\apj] {10.1088/0004-637X/721/2/1014},
  \href {http://cdsads.u-strasbg.fr/abs/2010ApJ...721.1014M} {721, 1014}

\bibitem[\protect\citeauthoryear{{Mainzer} et~al.,}{{Mainzer}
  et~al.}{2014}]{2014ApJ...792...30M}
{Mainzer} A.,  et~al., 2014, \mn@doi [\apj] {10.1088/0004-637X/792/1/30}, \href
  {http://adsabs.harvard.edu/abs/2014ApJ...792...30M} {792, 30}

\bibitem[\protect\citeauthoryear{{Mandal} et~al.,}{{Mandal}
  et~al.}{2018}]{2018MNRAS.475.5330M}
{Mandal} A.~K.,  et~al., 2018, \mn@doi [\mnras] {10.1093/mnras/sty200}, \href
  {http://adsabs.harvard.edu/abs/2018MNRAS.475.5330M} {475, 5330}

\bibitem[\protect\citeauthoryear{{Massaro}, {D'Abrusco}, {Ajello}, {Grindlay}
  \& {Smith}}{{Massaro} et~al.}{2011}]{2011ApJ...740L..48M}
{Massaro} F.,  {D'Abrusco} R.,  {Ajello} M.,  {Grindlay} J.~E.,   {Smith}
  H.~A.,  2011, \mn@doi [\apjl] {10.1088/2041-8205/740/2/L48}, \href
  {http://adsabs.harvard.edu/abs/2011ApJ...740L..48M} {740, L48}

\bibitem[\protect\citeauthoryear{{Massaro}, {D'Abrusco}, {Tosti}, {Ajello},
  {Gasparrini}, {Grindlay}  \& {Smith}}{{Massaro}
  et~al.}{2012}]{2012ApJ...750..138M}
{Massaro} F.,  {D'Abrusco} R.,  {Tosti} G.,  {Ajello} M.,  {Gasparrini} D.,
  {Grindlay} J.~E.,   {Smith} H.~A.,  2012, \mn@doi [\apj]
  {10.1088/0004-637X/750/2/138}, \href
  {http://adsabs.harvard.edu/abs/2012ApJ...750..138M} {750, 138}

\bibitem[\protect\citeauthoryear{{Maune}, {Miller}  \& {Eggen}}{{Maune}
  et~al.}{2013}]{2013ApJ...762..124M}
{Maune} J.~D.,  {Miller} H.~R.,   {Eggen} J.~R.,  2013, \mn@doi [\apj]
  {10.1088/0004-637X/762/2/124}, \href
  {http://adsabs.harvard.edu/abs/2013ApJ...762..124M} {762, 124}

\bibitem[\protect\citeauthoryear{{Meusinger}, {Hinze}  \& {de
  Hoon}}{{Meusinger} et~al.}{2011}]{2011A&A...525A..37M}
{Meusinger} H.,  {Hinze} A.,   {de Hoon} A.,  2011, \mn@doi [\aap]
  {10.1051/0004-6361/201015520}, \href
  {http://cdsads.u-strasbg.fr/abs/2011A%26A...525A..37M} {525, A37}

\bibitem[\protect\citeauthoryear{{Osterbrock} \& {Pogge}}{{Osterbrock} \&
  {Pogge}}{1985}]{1985ApJ...297..166O}
{Osterbrock} D.~E.,  {Pogge} R.~W.,  1985, \mn@doi [\apj] {10.1086/163513},
  \href {http://adsabs.harvard.edu/abs/1985ApJ...297..166O} {297, 166}

\bibitem[\protect\citeauthoryear{{Paliya}, {Stalin}, {Kumar}, {Kumar}, {Bhatt},
  {Pandey}  \& {Yadav}}{{Paliya} et~al.}{2013a}]{2013MNRAS.428.2450P}
{Paliya} V.~S.,  {Stalin} C.~S.,  {Kumar} B.,  {Kumar} B.,  {Bhatt} V.~K.,
  {Pandey} S.~B.,   {Yadav} R.~K.~S.,  2013a, \mn@doi [\mnras]
  {10.1093/mnras/sts217}, \href
  {http://adsabs.harvard.edu/abs/2013MNRAS.428.2450P} {428, 2450}

\bibitem[\protect\citeauthoryear{{Paliya}, {Stalin}, {Shukla}  \&
  {Sahayanathan}}{{Paliya} et~al.}{2013b}]{2013ApJ...768...52P}
{Paliya} V.~S.,  {Stalin} C.~S.,  {Shukla} A.,   {Sahayanathan} S.,  2013b,
  \mn@doi [\apj] {10.1088/0004-637X/768/1/52}, \href
  {http://adsabs.harvard.edu/abs/2013ApJ...768...52P} {768, 52}

\bibitem[\protect\citeauthoryear{{Paliya}, {Ajello}, {Rakshit}, {Mandal},
  {Stalin}, {Kaur}  \& {Hartmann}}{{Paliya} et~al.}{2018}]{2018ApJ...853L...2P}
{Paliya} V.~S.,  {Ajello} M.,  {Rakshit} S.,  {Mandal} A.~K.,  {Stalin} C.~S.,
  {Kaur} A.,   {Hartmann} D.,  2018, \mn@doi [\apjl]
  {10.3847/2041-8213/aaa5ab}, \href
  {http://adsabs.harvard.edu/abs/2018ApJ...853L...2P} {853, L2}

\bibitem[\protect\citeauthoryear{{Polimera}, {Sarajedini}, {Ashby}, {Willner}
  \& {Fazio}}{{Polimera} et~al.}{2018}]{2018MNRAS.476.1111P}
{Polimera} M.,  {Sarajedini} V.,  {Ashby} M.~L.~N.,  {Willner} S.~P.,   {Fazio}
  G.~G.,  2018, \mn@doi [\mnras] {10.1093/mnras/sty164}, \href
  {http://adsabs.harvard.edu/abs/2018MNRAS.476.1111P} {476, 1111}

\bibitem[\protect\citeauthoryear{{Pounds}, {Done}  \& {Osborne}}{{Pounds}
  et~al.}{1995}]{1995MNRAS.277L...5P}
{Pounds} K.~A.,  {Done} C.,   {Osborne} J.~P.,  1995, \mn@doi [\mnras]
  {10.1093/mnras/277.1.L5}, \href
  {http://adsabs.harvard.edu/abs/1995MNRAS.277L...5P} {277, L5}

\bibitem[\protect\citeauthoryear{{Rakshit} \& {Stalin}}{{Rakshit} \&
  {Stalin}}{2017}]{2017ApJ...842...96R}
{Rakshit} S.,  {Stalin} C.~S.,  2017, \mn@doi [\apj]
  {10.3847/1538-4357/aa72f4}, \href
  {http://adsabs.harvard.edu/abs/2017ApJ...842...96R} {842, 96}

\bibitem[\protect\citeauthoryear{{Rakshit}, {Petrov}, {Meilland}  \&
  {H{\"o}nig}}{{Rakshit} et~al.}{2015}]{2015MNRAS.447.2420R}
{Rakshit} S.,  {Petrov} R.~G.,  {Meilland} A.,   {H{\"o}nig} S.~F.,  2015,
  \mn@doi [\mnras] {10.1093/mnras/stu2613}, \href
  {http://adsabs.harvard.edu/abs/2015MNRAS.447.2420R} {447, 2420}

\bibitem[\protect\citeauthoryear{{Rakshit}, {Stalin}, {Chand}  \&
  {Zhang}}{{Rakshit} et~al.}{2017a}]{2017ApJS..229...39R}
{Rakshit} S.,  {Stalin} C.~S.,  {Chand} H.,   {Zhang} X.-G.,  2017a, \mn@doi
  [\apjs] {10.3847/1538-4365/aa6971}, \href
  {http://adsabs.harvard.edu/abs/2017ApJS..229...39R} {229, 39}

\bibitem[\protect\citeauthoryear{{Rakshit}, {Stalin}, {Muneer}, {Neha}  \&
  {Paliya}}{{Rakshit} et~al.}{2017b}]{2017ApJ...835..275R}
{Rakshit} S.,  {Stalin} C.~S.,  {Muneer} S.,  {Neha} S.,   {Paliya} V.~S.,
  2017b, \mn@doi [\apj] {10.3847/1538-4357/835/2/275}, \href
  {http://adsabs.harvard.edu/abs/2017ApJ...835..275R} {835, 275}

\bibitem[\protect\citeauthoryear{{Rees}}{{Rees}}{1984}]{1984ARA&A..22..471R}
{Rees} M.~J.,  1984, \mn@doi [\araa] {10.1146/annurev.aa.22.090184.002351},
  \href {http://cdsads.u-strasbg.fr/abs/1984ARA%26A..22..471R} {22, 471}

\bibitem[\protect\citeauthoryear{{Romero}, {Cellone}  \& {Combi}}{{Romero}
  et~al.}{1999}]{1999A&AS..135..477R}
{Romero} G.~E.,  {Cellone} S.~A.,   {Combi} J.~A.,  1999, \mn@doi [\aaps]
  {10.1051/aas:1999184}, \href
  {http://adsabs.harvard.edu/abs/1999A%26AS..135..477R} {135, 477}

\bibitem[\protect\citeauthoryear{{Rumbaugh} et~al.,}{{Rumbaugh}
  et~al.}{2018}]{2018ApJ...854..160R}
{Rumbaugh} N.,  et~al., 2018, \mn@doi [\apj] {10.3847/1538-4357/aaa9b6}, \href
  {http://adsabs.harvard.edu/abs/2018ApJ...854..160R} {854, 160}

\bibitem[\protect\citeauthoryear{{Sesar} et~al.,}{{Sesar}
  et~al.}{2007}]{2007AJ....134.2236S}
{Sesar} B.,  et~al., 2007, \mn@doi [\aj] {10.1086/521819}, \href
  {http://cdsads.u-strasbg.fr/abs/2007AJ....134.2236S} {134, 2236}

\bibitem[\protect\citeauthoryear{{Shakura} \& {Sunyaev}}{{Shakura} \&
  {Sunyaev}}{1973}]{1973A&A....24..337S}
{Shakura} N.~I.,  {Sunyaev} R.~A.,  1973, \aap, \href
  {http://adsabs.harvard.edu/abs/1973A%26A....24..337S} {24, 337}

\bibitem[\protect\citeauthoryear{{Stern} et~al.,}{{Stern}
  et~al.}{2012}]{2012ApJ...753...30S}
{Stern} D.,  et~al., 2012, \mn@doi [\apj] {10.1088/0004-637X/753/1/30}, \href
  {http://adsabs.harvard.edu/abs/2012ApJ...753...30S} {753, 30}

\bibitem[\protect\citeauthoryear{{Suganuma} et~al.,}{{Suganuma}
  et~al.}{2006}]{2006ApJ...639...46S}
{Suganuma} M.,  et~al., 2006, \mn@doi [\apj] {10.1086/499326}, \href
  {http://adsabs.harvard.edu/abs/2006ApJ...639...46S} {639, 46}

\bibitem[\protect\citeauthoryear{{Ulrich}, {Maraschi}  \& {Urry}}{{Ulrich}
  et~al.}{1997}]{1997ARA&A..35..445U}
{Ulrich} M.-H.,  {Maraschi} L.,   {Urry} C.~M.,  1997, \mn@doi [\araa]
  {10.1146/annurev.astro.35.1.445}, \href
  {http://cdsads.u-strasbg.fr/abs/1997ARA%26A..35..445U} {35, 445}

\bibitem[\protect\citeauthoryear{{Urry} \& {Padovani}}{{Urry} \&
  {Padovani}}{1995}]{1995PASP..107..803U}
{Urry} C.~M.,  {Padovani} P.,  1995, \mn@doi [\pasp] {10.1086/133630}, \href
  {http://cdsads.u-strasbg.fr/abs/1995PASP..107..803U} {107, 803}

\bibitem[\protect\citeauthoryear{{Vanden Berk} et~al.,}{{Vanden Berk}
  et~al.}{2004}]{2004ApJ...601..692V}
{Vanden Berk} D.~E.,  et~al., 2004, \mn@doi [\apj] {10.1086/380563}, \href
  {http://cdsads.u-strasbg.fr/abs/2004ApJ...601..692V} {601, 692}

\bibitem[\protect\citeauthoryear{{Wagner} \& {Witzel}}{{Wagner} \&
  {Witzel}}{1995}]{1995ARA&A..33..163W}
{Wagner} S.~J.,  {Witzel} A.,  1995, \mn@doi [\araa]
  {10.1146/annurev.aa.33.090195.001115}, \href
  {http://cdsads.u-strasbg.fr/abs/1995ARA%26A..33..163W} {33, 163}

\bibitem[\protect\citeauthoryear{{Wang}, {Brinkmann}  \& {Bergeron}}{{Wang}
  et~al.}{1996}]{1996A&A...309...81W}
{Wang} T.,  {Brinkmann} W.,   {Bergeron} J.,  1996, \aap, \href
  {http://cdsads.u-strasbg.fr/abs/1996A%26A...309...81W} {309, 81}

\bibitem[\protect\citeauthoryear{{Wright} et~al.,}{{Wright}
  et~al.}{2010}]{2010AJ....140.1868W}
{Wright} E.~L.,  et~al., 2010, \mn@doi [\aj] {10.1088/0004-6256/140/6/1868},
  \href {http://adsabs.harvard.edu/abs/2010AJ....140.1868W} {140, 1868}

\bibitem[\protect\citeauthoryear{{Xu}, {Komossa}, {Zhou}, {Lu}, {Li}, {Grupe},
  {Wang}  \& {Yuan}}{{Xu} et~al.}{2012}]{2012AJ....143...83X}
{Xu} D.,  {Komossa} S.,  {Zhou} H.,  {Lu} H.,  {Li} C.,  {Grupe} D.,  {Wang}
  J.,   {Yuan} W.,  2012, \mn@doi [\aj] {10.1088/0004-6256/143/4/83}, \href
  {http://cdsads.u-strasbg.fr/abs/2012AJ....143...83X} {143, 83}

\bibitem[\protect\citeauthoryear{{Yang} et~al.,}{{Yang}
  et~al.}{2018}]{2018ApJ...862..109Y}
{Yang} Q.,  et~al., 2018, \mn@doi [\apj] {10.3847/1538-4357/aaca3a}, \href
  {http://cdsads.u-strasbg.fr/abs/2018ApJ...862..109Y} {862, 109}

\bibitem[\protect\citeauthoryear{{di Clemente}, {Giallongo}, {Natali},
  {Trevese}  \& {Vagnetti}}{{di Clemente} et~al.}{1996}]{1996ApJ...463..466D}
{di Clemente} A.,  {Giallongo} E.,  {Natali} G.,  {Trevese} D.,   {Vagnetti}
  F.,  1996, \mn@doi [\apj] {10.1086/177261}, \href
  {http://cdsads.u-strasbg.fr/abs/1996ApJ...463..466D} {463, 466}

\makeatother
\end{thebibliography}


\label{lastpage}
\end{document}